\documentclass[11pt]{article}
\usepackage{mathrsfs}
\usepackage{amssymb}
\usepackage{color}
\usepackage{amsmath}
\usepackage{ascmac}
\usepackage{amsthm}
\usepackage[dvips]{graphicx}
\usepackage{booktabs}

\newtheorem{thm}{Theorem}

\newtheorem{cor}{Corollary}

\def\be{{\beta}}
\def\ga{{\gamma}}

\def\ep{{\varepsilon}}
\def\la{{\lambda}}
\def\si{{\sigma}}

\def\th{{\theta}}

\def\ka{{\kappa}}

\def\bbe{{\text{\boldmath $\beta$}}}
\def\bga{{\text{\boldmath $\gamma$}}}
\def\bta{{\text{\boldmath $\eta$}}}
\def\bde{{\text{\boldmath $\delta$}}}
\def\bep{{\text{\boldmath $\epsilon$}}}

\def\bth{{\text{\boldmath $\theta$}}}

\def\bka{{\text{\boldmath $\kappa$}}}

\def\bpsi{{\text{\boldmath $\psi$}}}
\def\bphi{{\text{\boldmath $\phi$}}}

\def\beh{{\widehat \be}}

\def\etah{{\widehat \eta}}
\def\gah{{\widehat \ga}}

\def\thh{{\widehat \th}}

\def\sih{{\widehat \si}}

\def\tah{{\widehat \tau}}
\def\muh{{\widehat \mu}}
\def\lah{{\hat \la}}

\def\bbeh{{\widehat \bbe}}
\def\bgah{{\widehat \bga}}
\def\bphih{{\widehat \bphi}}

\def\bthh{{\widehat \bth}}

\def\bkah{\widehat{\bka}}

\def\bbet{{\widetilde \bbe}}

\def\bgat{{\widetilde \bga}}

\def\tat{{\widetilde \tau}}
\def\Si{{\Sigma}}
\def\bSi{{\text{\boldmath $\Si$}}}
\def\0{{\text{\boldmath $0$}}}
\def\a{{\text{\boldmath $a$}}}
\def\b{{\text{\boldmath $b$}}}
\def\c{{\text{\boldmath $c$}}}

\def\t{{\text{\boldmath $t$}}}
\def\u{{\text{\boldmath $u$}}}
\def\v{{\text{\boldmath $v$}}}

\def\x{{\text{\boldmath $x$}}}
\def\y{{\text{\boldmath $y$}}}
\def\z{{\text{\boldmath $z$}}}

\def\E{{\text{\boldmath $E$}}}
\def\F{{\text{\boldmath $F$}}}

\def\I{{\text{\boldmath $I$}}}
\def\J{{\text{\boldmath $J$}}}

\def\M{{\text{\boldmath $M$}}}

\def\T{{\text{\boldmath $T$}}}

\def\V{{\text{\boldmath $V$}}}
\def\W{{\text{\boldmath $W$}}}
\def\X{{\text{\boldmath $X$}}}

\def\Z{{\text{\boldmath $Z$}}}

\def\diag{{\rm diag}}
\def\tr{{\rm tr}}
\def\1{{\text{\boldmath $1$}}}

\def\barx{\bar{\x}}

\def\bary{\bar{y}}
\def\barz{\bar{\z}}
\def\bOm{{\text{\boldmath $\Omega$}}}

\def\Obbeh{\bbeh_{\rm OLS}}
\def\mut{{\widetilde \mu}}
\def\lah{{\widehat \la}}
\def\kah{{\widehat \ka}}
\def\barep{\bar{\ep}}
\def\siij{\si_{ij}}
\def\siih{\si_{ih}}

\def\Var{{\rm Var}}

\title{Heteroscedastic Nested Error Regression Models\\
 with Variance Functions}

\author{
Shonosuke Sugasawa\thanks{The Institute of Statistical Mathematics, Email: sugasawa@ism.ac.jp}
 \ and Tatsuya Kubokawa\thanks{Faculty of Economics, University of Tokyo}
}

\date{}

\begin{document}

\maketitle

\begin{abstract}
The nested error regression model is a useful tool for analyzing clustered (grouped) data, and is especially used in small area estimation.
The classical nested error regression model assumes normality of random effects and error terms, and homoscedastic variances.
However, these assumptions are often violated in real applications and more flexible models are required. 
This article proposes a nested error regression model with heteroscedastic variances, where the normality for the underlying distributions is not assumed. 
We propose the structure of heteroscedastic variances by using some specified variance functions and some covariates with unknown parameters.
Under the setting, we construct the moment-type estimators of model parameters and some asymptotic properties including asymptotic biases and variances are derived.
For predicting linear quantities including random effects, we suggest the empirical best linear unbiased predictors and the second-order unbiased estimators of mean squared errors are derived in the closed form.
We investigate the proposed method with simulation and empirical studies.

\ \\
{\it keywords and phrases}:
empirical best linear unbiased predictor;
heteroscedastic variance;
mean squared error;
nested error regression;
small area estimation; 
variance function
\end{abstract}

\section{Introduction}
Linear mixed models and the model-based estimators including empirical Bayes (EB) estimator or empirical best linear unbiased predictor (EBLUP) have been studied quite extensively in the literature from both theoretical and applied points of view.
Of these, the small area estimation (SAE) is an important application, and methods for SAE have received much attention in recent years due to growing demand for reliable small area estimates.
For a good review on this topic, see Ghosh and Rao (1994), Rao and Molina (2015), Datta and Ghosh (2012) and Pfeffermann (2014).
The linear mixed models used for SAE are the Fay-Herriot model suggested by Fay and Herriot (1979) for area-level data and the nested error regression (NER) models given in Battese, Harter and Fuller (1988) for unit-level data.
Especially, the NER model has been used in application of not only SAE but also biological experiments and econometric analysis.
In the NER model, a cluster-specific variation is added to explain the correlation among observations within clusters besides the noise, which allow the analysis to ^^ borrow strength' from other clusters.
The resulting estimators, such as EB or EBLUP, for small-cluster means or subject-specific values  provide reliable estimates with higher precisions than direct estimates like sample means.

In the NER model with $m$ small-clusters, let $(y_{i1}, \x_{i1}),  \ldots, (y_{i n_i}, \x_{i n_i})$ be $n_i$ individual observations from the $i$-th cluster for $i=1, \ldots, m$, where $\x_{ij}$ is a $p$-dimensional known vector of covariates.
The normal NER model proposed by Battese, et al. (1998) is given by
$$
y_{ij} = \x_{ij}'\bbe + v_i + \ep_{ij}, \quad i=1, \ldots, m, \quad j=1, \ldots, n_i,
$$
where $v_i$ and $\ep_{ij}$ denote the random effect and samping error, respectively, and they are mutually independently distributed as $v_i\sim N(0, \tau^2)$ and $\ep_{ij}\sim N(0, \si^2)$. 
The mean of $y_{ij}$ is $\x_{ij}'\bbe$ for regression coefficients $\bbe$, and the variance of $y_{ij}$ is decomposed as $\Var(y_{ij})=\tau^2 + \si^2$, which is the same for all the clusters.
However, Jiang and Nguyen (2012) illustrated that the within-cluster sample variances change dramatically from cluster to cluster for the data given in Battese, et al. (1988).
Then, Jiang and Nguyen (2012) proposed the heteroscedastic nested error regression (HNER) model with the setup that variance $\Var(y_{ij})$ is proportional to $\si_i^2$, namely $\Var(y_{ij})= (\la + 1)\si_i^2$.
This is equivalent to the assumption that $\Var(v_i)=\la\si_i^2$ and $\Var(\ep_{ij})=\si_i^2$.
Under this setup, Jiang and Nguyen (2012) assumed normality for $v_i$ and $\ep_{ij}$ and demonstrated the quite interesting result that the maximum likelihood (ML) estimators of $\bbe$ and $\la$ are consistent for large $m$, which implies that the resulting EB estimator is asymptotically equivalent to the Bayes estimator.
Thorough simulation studies, Jiang and Nguyen (2012) also showed that that the EBLUP from HNER model can improve the prediction accuracy over that from NER model when the data is generated from HNER model.
However, there is no consistent estimator for the heteroscedastic variance $\si_i^2$ because of finiteness of $n_i$, and the mean squared error (MSE) of the EBLUP cannot be estimated consistently since it depends on $\si_i^2$.
To fix the inconsistent estimation of $\si_i^2$, recently, Kubokawa, Sugasawa, Ghosh and Chaudhuri (2016) proposed the hierarchical model such that $\si_i^2$'s are random variables and $\si_i^{-2}$ has a gamma distribution.
The same dispersion structure was used in Maiti, Ren and Sinha (2014) who applied this hierarchical structure to the Fay-Herriot model (Fay and Herriot, 1979) with statistics for estimating $\si_i^2$.
Kubokawa, et al. (2016) proposed the ML estimators of model parameters including the shape and scale parameters in dispersion distribution of $\si_i^2$.
They also showed the consistency of the model parameters and constructed the second-order unbiased mean squared errors of MSE by using parametric bootstrap.

While these two HNER models are useful for analyzing unit-level data with heteroscedastic variances, the serious drawback of the two models is that both require normality assumption for random effects and error terms, which are not necessary satisfied in real application.
Hence, the purpose of this paper is to address the issue of relaxing assumptions of classical normal NER models toward two directions: heteroscedasticity of variances and non-normality of underlying distributions.

In real data analysis, we often encounter the situation where the sampling variance $\Var(\ep_{ij})$ is affected by the covariate $\x_{ij}$.
In such case, the variance function is a useful tool for describing its relationship.
Variance function estimation has been studied in the literature in the framework of heteroscedastic nonparametric regression.  
For example, see Cook and Weisberg (1983), Hall and Carroll (1989), Muller and Stadtmuller (1987, 1993) and Ruppert, Wand, Holst and Hossjer (1997).
Thus, in this paper, we propose the use of the technique to introduce the heteroscedastic variances into the NER model without assuming normality of underlying distributions.
The variance structure we consider is $\Var(y_{ij})= \tau^2 + \siij^2$, namely, the setup means that the sampling error $\ep_{ij}$ has heteroscedastic variance $\Var(\ep_{ij})=\siij^2$.
Then we suggest the variance function model given by $\siij^2 = \si^2(\z_{ij}'\bga)$, where the details are explained in Section \ref{sec:model}.
In terms of modeling the heteroscedastic variances with covariates, the generalized linear mixed models (Jiang, 2006) are also the useful tool.
The small area models using generalized linear mixed models are investigated in Ghosh, Natarajan, Stroud and Carlin (1998). 
However, the generalized linear mixed model requires strong parametric assumption compared to the heteroscedastic model without assuming underlying distributions proposed in this paper.
Hence, the generalized linear mixed model seems still restrictive while it is an attractive method for modelling heteroscedasticity in variances.

In this paper, we propose flexible and tractable HNER models without assuming normality for either $v_i$ nor $\ep_{ij}$.  
The advantage of the proposed model is that the MSE of the EB or EBLUP and its unbiased estimator are derived analytically in closed forms up to second-order without assuming normality for $v_i$ and $\ep_{ij}$.
Nonparametric approach to SAE has been studied by Jiang, Lahiri and Wan (2002), Hall and Maiti (2006), Lohr and Rao (2009) and others.
Most estimators of the MSE have been given by numerical methods such as Jackknife and bootstrap methods except for Lahiri and Rao (1995), who provided an analytical second-order unbiased estimator of the MSE in the Fay-Heriot model.
Hall and Maiti (2006) developed a moment matching bootstrap method for nonparametric estimation of MSE in nested error regression models.
The suggested method is actually convenient but it requires bootstrap replication and has computational burden.  
In this paper, without assuming the normality, we derive a closed expression for a second-order unbiased estimator of the MSE using second-order biases and variances of estimators of the model parameters.
Thus our MSE estimator does not require any resampling method and is convenient in practical use.
Also our MSE estimator can be regarded as a generalization of the robust MSE estimator given in Lahiri and Rao (1995).

The paper is organized as follows:
A setup of the proposed HNER model and estimation strategy with asymptotic properties are given in Section \ref{sec:model}.
In Section \ref{sec:mse}, we obtain the EBLUP and the second-order approximation of the MSE.
Further, we provide the second-order unbiased estimators of MSE by the analytical calculation.
In Section \ref{sec:sim}, we investigate the performance of the proposed procedures through simulation and empirical studies.
All the technical proofs are given in the Appendix.

\section{HNER Models with Variance Functions}\label{sec:model}

\subsection{Model settings}\label{sec:setting}

Suppose that there are $m$ small clusters, and let $(y_{i1}, \x_{i1}), \ldots, (y_{i n_i}, \x_{i n_i})$ be the pairs of $n_i$ observations from the $i$-th cluster, where $\x_{ij}$ is a $p$-dimensional known vector of covariates.
We consider the heteroscedastic nested error regression model
\begin{equation}\label{model}
y_{ij}=\x_{ij}'\bbe+v_i+\ep_{ij}, \ \ j=1,\ldots,n_i, \ \ i=1,\ldots,m,
\end{equation}
where $\bbe$ is a $p$-dimensional unknown vector of regression coefficients, and $v_i$ and $\ep_{ij}$ are mutually independent random variables with mean zero and variances $\Var(v_i)=\tau^2$ and $\Var(\ep_{ij})=\siij^2$, which are denoted by 
\begin{equation}\label{dist}
v_i\sim (0,\tau^2)\quad \text{and}\quad \ep_{ij}\sim (0,\siij^2).
\end{equation}
It is noted that no specific distributions are assumed for $v_i$ and $\ep_{ij}$.
It is assumed that the heteroscedastic variance $\siij^2$ of $\ep_{ij}$ is given by
\begin{equation}
\siij^2=\si^2(\z_{ij}'\bga), \quad i=1, \ldots, m,
\label{VF}
\end{equation}
where $\z_{ij}$ is a $q$-dimensional known vector given for each cluster, and $\bga$ is a $q$-dimensional unknown vector.
The variance function $\si^2(\cdot)$ is a known (user specified) function whose range is nonnegative.
Some examples of the variance function are given below.
The model parameters are $\bbe$, $\tau^2$ and $\bga$, and the total number of the model parameters is $p+q+1$.

Let $\y_i=(y_{i1},\ldots,y_{in_i})'$,  $\X_i=(\x_{i1},\ldots,\x_{in_i})'$ and $\bep_i=(\ep_{i1},\ldots,\ep_{in_i})'$.
Then the model (\ref{model}) is expressed in a vector form as
$$
\y_i=\X_i\bbe+v_i\1_{n_i}+\bep_{i}, \ \ \ \ i=1,\ldots,m,
$$
where $\1_{n}$ is an $n\times 1$ vector with all elements equal to one, and the covariance matrix of $\bep_i$ is
$$
\bSi_i=\Var(\y_i)=\tau^2\J_{n_i}+\W_i,
$$
for $\J_{n_i}=\1_{n_i}\1_{n_i}'$ and $\W_i=\diag(\si_{i1}^{2},\ldots,\si_{in_i}^{2})$.
It is noted that the inverse of $\bSi_i$ is expressed as
$$
\bSi_i^{-1}=\W_i^{-1}\left(\I_{n_i}-\frac{\tau^2\J_{n_i}\W_i^{-1}}{1+\tau^2\sum_{j=1}^{n_i}\siij^{-2}}\right),
$$
where $\W_i^{-1}=\diag(\si_{i1}^{-2},\ldots,\si_{in_i}^{-2})$.
Further, let $\y=(\y_1',\ldots,\y_m')'$, $\X=(\X_1',\ldots,\X_m')'$, $\bep=(\bep_1',\ldots,\bep_m')'$ and $\v=(v_1\1_{n_1}',\ldots,v_m\1_{n_m}')'$.
Then, the matricial form of (\ref{model}) is written as $\y=\X\bbe+\v+\bep$, where $\Var(\y)=\bSi=\text{block diag}(\bSi_1,\ldots,\bSi_m)$.
Now we give three examples of the variance function $\si^2(\z_{ij}'\bga)$ in (\ref{VF}).

\begin{itemize}
\item[(a)]
In the case that the dispersion of the sampling error is proportional to the mean, it is reasonable to put $\z_{ij}=\x_{(s)ij}$ and $\si^2(\x_{(s)ij}'\bga)=(\x_{(s)ij}'\bga)^2$ for a sub-vector $\x_{(s)ij}$ of the covariate $\x_{ij}$.
For identifiability of $\bga$, we restrict $\ga_1>0$.

\item[(b)]
Consider the case that $m$ clusters are decomposed into $q$ homogeneous groups $S_1,\ldots,S_q$ with $\{1,\ldots,m\}=S_1\cup\ldots\cup S_q$.
Then, we put
$$
\z_{ij}=\left(1_{\{i\in S_1\}},\ldots,1_{\{i\in S_q\}}\right)',
$$
which implies that 
$$
\siij^2 = \ga_t^2 \quad \text{for}\quad i\in S_t.
$$
Note that $\Var(y_{ij})=\tau^2+\ga_t^2$ for $i\in S_t$.
Thus, the models assumes that the $m$ clusters are divided into known $q$ groups with their variance are equal over the same groups.
Jiang and Nguyen (2012) used a similar setting and argued that the unbiased estimator of the heteroscedastic variance is consistent when $|S_k|\to\infty, k=1,\ldots,q$ as $m\to\infty$, where $|S_k|$ denotes the number of elements in $S_k$.

\item[(c)]
Log linear functions of variance were treated in Cook and Weisberg (1983) and others.
That is, $\log \siij^2$ is a linear function, and $\siij^2$ is written as $\si^2(\z_{ij}'\bga)=\exp( \z_{ij}'\bga )$.
Similarly to (a), we put $\z_{ij}=\x_{(s)ij}$.
\end{itemize}

For the above two cases (a) and (b), we have $\si^2(x)=x^2$, while the case (c) corresponds to $\log \{\si^2(x)\}= x$.
In simulation and empirical studies in Section \ref{sec:sim}, we use the log-linear variance model.
As given in the subsequent section, we show consistency and asymptotic expression of estimators for $\bga$ as well as $\bbe$ and $\tau^2$.

\subsection{Estimation}

We here provide estimators of the model parameters $\bbe$, $\tau^2$ and $\bga$.
When values of $\bga$ and $\tau^2$ are given, the vector $\bbe$ of regression coefficients is estimated by the generalized least squares (GLS) estimator
\begin{equation}\label{bbe}
\begin{split}
\bbet&=\bbet(\tau^2,\bga)=(\X'\bSi^{-1}\X)^{-1}\X'\bSi^{-1}\y=\left(\sum_{i=1}^m\X_i'\bSi_i^{-1}\X_i\right)^{-1}\sum_{i=1}^m\X_i'\bSi^{-1}_i\y_i.
\end{split}
\end{equation}
This is not a feasible form since $\bga$ and $\tau^2$ are unknown.
When estimators $\tah^2$ and $\bgah$ are used for $\tau^2$ and $\bga$, we get the feasible estimator $\bbeh=\bbet(\tah^2, \bgah)$ by replacing $\tau^2$ and $\bga$ in $\bbet$ with their estimators.

Concerning estimation of $\tau^2$, we use the second moment of observations $y_{ij}$'s.
From model (\ref{model}), it is seen that
\begin{equation}\label{mom2}
E\left[(y_{ij}-\x_{ij}'\bbe)^2\right]=\tau^2+\si^2(\z_{ij}'\bga).
\end{equation}
Based on the ordinary least squares (OLS) estimator $\bbeh_{\rm OLS}=(\X'\X)^{-1}\X'\y$, a moment estimator of $\tau^2$ is given by
\begin{equation}\label{tau}
\tah^2=\frac1N\sum_{i=1}^m\sum_{j=1}^{n_i}\left\{(y_{ij}-\x_{ij}'\bbeh_{\rm OLS})^2-\si^2(\z_{ij}'\bga)\right\},
\end{equation}
with substituting estimator $\bgah$ into $\bga$, where $N=\sum_{i=1}^{m}n_i$.

For estimation of $\bga$, we consider the within difference in each cluster.
Let $\bary_i$ be the sample mean in the $i$-th cluster, namely $\bary_i=n_i^{-1}\sum_{j=1}^{n_i}y_{ij}$.  
It is noted that for $\bar{\ep}_i=n_i^{-1}\sum_{j=1}^{n_i}\ep_{ij}$,
$$
y_{ij}-\bary_i=(\x_{ij}-\barx_i)'\bbe+(\ep_{ij}-\bar{\ep}_i),
$$ 
which dose not include the term of $v_i$.
Then it is seen that
$$
E\left[\left\{y_{ij}-\bary_i-(\x_{ij}-\barx_i)'\bbe\right\}^2\right]=\left(1-2n_i^{-1}\right)\si^2(\z_{ij}'\bga)+n_i^{-2}\sum_{h=1}^{n_i}\si^2(\z_{ih}'\bga),
$$
which motivates us to estimate $\bga$ by solving the following estimating equation given by
$$
\frac1N\sum_{i=1}^m\sum_{j=1}^{n_i}\left[\left\{y_{ij}-\bary_i-(\x_{ij}-\barx_i)'\Obbeh\right\}^2-\left(1-2n_i^{-1}\right)\si^2(\z_{ij}'\bga)-n_i^{-2}\sum_{h=1}^{n_i}\si^2(\z_{ih}'\bga)\right]\z_{ij}=\0,
$$
which is equivalent to
\begin{equation}\label{bga}
\frac1N\sum_{i=1}^m\sum_{j=1}^{n_i}\left[\left\{y_{ij}-\bary_i-(\x_{ij}-\barx_i)'\Obbeh\right\}^2\z_{ij}-\si^2(\z_{ij}'\bga)(\z_{ij}-2n_i^{-1}\z_{ij}+n_i^{-1}\barz_i)\right]=\0
\end{equation}
where $\barz_i=n_i^{-1}\sum_{j=1}^{n_i}\z_{ij}$.
It is noted that, in the homoscedastic case with $\si^2(\z_{ij}'\bga)=\delta^2$, the estimators of $\delta^2$ and $\tau^2$ reduce to the estimators identical to the Prasad-Rao estimator (Prasad and Rao, 1990) up to the constant factor.

Note that the function given in the left side of (\ref{bga}) does not depend on $\bbe$ and $\tau^2$ and the estimator of $\tau^2$ does not depend on $\bbe$ but on $\bga$.
These suggest the simple algorithm for calculating the estimates of the model parameters:
We first obtain the estimate $\bgah$ of $\bga$ by solving (\ref{bga}), and then we get the estimate $\tah^2$ from (\ref{tau}) with $\bga=\bgah$.
Finally we have the GLS estimate $\bbeh$ with substituting  $\bgah$ and $\tah^2$ in (\ref{bbe}).

\subsection{Large sample properties}

In this section, we provide large sample properties of the estimators given in the previous subsection when the number of clusters $m$ goes to infinity, but $n_i$'s are still bounded.
To establish asymptotic results, we assume the following conditions under $m\to\infty$.

\vspace{0.3cm}
\noindent{\bf Assumption (A)}
\begin{itemize}
\item[(A1)]
There exist bounded values $\underline{n}$ and $\overline{n}$ such that $\underline{n}\leq n_i\leq \overline{n}$ for $i=1,\ldots,m$. 
The dimensions $p$ and $q$ are bounded, namely $p,q=O(1)$. 
The number of clusters with one observation, namely $n_i=1$, is bounded.

\item[(A2)]
The variance function $\si^2(\cdot)$ is twice differentiable and its derivatives are denoted by $(\si^2)^{(1)}(\cdot)$ and $(\si^2)^{(2)}(\cdot)$, respectively.

\item[(A3)]
The following matrices converge to non-singular matrices:
$$
m^{-1}\sum_{i=1}^m\sum_{j=1}^{n_i}\z_{ij}\z_{ij}', \ \ \ m^{-1}\sum_{i=1}^m\sum_{j=1}^{n_i}(\si^2)^{(a_1)}(\z_{ij}'\bga)\z_{ij}\z_{ij}', \ \ \ m^{-1}\X'\bSi^{a_2}\X
$$
 for $a_1=1,2$ and $a_2=-1,0,1$.

\item[(A4)]
$E[|v_i|^{8+c}]<\infty$ and $E[|\ep_{ij}|^{8+c}]<\infty$ for $0<c<1$.

\item[(A5)]
For all $i$ and $j$, there exist $0<\underline{c_1},\overline{c_1}<\infty$ and bounded values $\underline{c_2}, \overline{c_2}$ such that $\underline{c_1}<\si^2(\z_{ij}'\bga)<\overline{c_1}$ and $\underline{c_2}<(\si^2)^{(k)}(\z_{ij}'\bga)<\overline{c_2}$ with $k=1,2$ on the neighborhood of the true values.

\end{itemize}

\medskip\noindent   
The conditions (A1) and (A3) are the standard assumptions in small area estimation.
The condition (A2) is also non-restrictive, and the typical variance functions $\si^2(x)=x^2$ and $\si^2(x)=\exp x$ obviously satisfy the assumption. 
The moment condition (A4) is used for deriving second-order approximation of MSE of the EBLUP discussed in Section \ref{sec:mse}, and it is satisfied by many continuous distributions, including normal, shifted gamma, Laplace and $t$-distribution with degrees of freedom larger than 9.
The three examples given in Section \ref{sec:setting} satisfy the condition (A5).

In what follows, we use the notations 
$$
\siij^2\equiv \si^2(\z_{ij}'\bga), \ \ \ \ \si_{ij(k)}^2\equiv (\si^2)^{(k)}(\z_{ij}'\bga), \ \ k=1,2
$$ 
for simplicity.
To derive asymptotic approximations of the estimators, we use the following notations in the $i$-th cluster:
\begin{align}
u_{1i}&=\frac{m}{N}\sum_{j=1}^{n_i}\left\{(y_{ij}-\x_{ij}'\bbe)^2-\siij^2-\tau^2\right\},\label{u1}  \\
\u_{2i}&=\frac{m}{N}\sum_{j=1}^{n_i}\left[\left\{y_{ij}-\bary_i-(\x_{ij}-\barx_i)'\bbe\right\}^2\z_{ij}-\siij^2(\z_{ij}-2n_i^{-1}\z_{ij}+n_i^{-1}\barz_i)\right],
\label{u2}
\end{align}
with
\begin{equation}\label{T}\begin{split}
\T_1(\bga)&=\sum_{k=1}^m\sum_{h=1}^{n_k}\si_{kh(1)}^2\z_{kh}, \ \ \ \T_2(\bga)=\left(\sum_{k=1}^m\sum_{j=1}^{n_k}\si_{kh(1)}^2(\z_{kh}-2n_k^{-1}\z_{kh}+n_k^{-1}\barz_k)\z_{kh}'\right)^{-1}.
\end{split}\end{equation}
Note that $\T_1(\bga)=O(m)$ and $\T_2(\bga)=O(m^{-1})$ under Assumption (A).
Then we obtain the asymptotically linear expression of the estimators.

\vspace{0.5cm}
\begin{thm}\label{thm:asymp}
Let $\bthh=(\bbeh',\bgah',\tah^2)'$ be the estimator of $\bth=(\bbe',\bga',\tau^2)'$.
Under Assumption {\rm (A)}, it holds that $\bthh-\bth=O_p(m^{-1/2})$ with the asymptotically linear expression 
$$
\bthh-\bth=\frac1m\sum_{i=1}^m((\bpsi_i^{\bbe})',(\bpsi_i^{\bga})',\psi_i^{\tau})'+o_p(m^{-1/2}),
$$
where
\begin{align*}
\bpsi_i^{\bbe}=m \left(\X'\bSi^{-1}\X\right)^{-1}\X_i\bSi_i^{-1}(\y_i-\X_i\bbe), \ \ \ \ 
\bpsi_i^{\bga}=N\T_2(\bga)\u_{2i}, \ \ \ \ 
\psi_i^{\tau}=u_{1i}-\T_1(\bga)'\T_2(\bga)\u_{2i}.\ \ \ \ \ 
\end{align*}
\end{thm}

\vspace{0.5cm}
From Theorem \ref{thm:asymp}, it follows that $m^{1/2}(\bthh-\bth)$ has an asymptotically normal distribution with mean vector $\0$ and covariance matrix $m\bOm$, where $\bOm$ is a $(p+q+1)\times (p+q+1)$ matrix partitioned as
\begin{align*}
m\bOm\equiv \left(\begin{array}{ccc}
m\bOm_{\bbe\bbe} & m\bOm_{\bbe\bga} & m\bOm_{\bbe\tau}\\
m\bOm_{\bbe\bga}' & m\bOm_{\bga\bga} & m\bOm_{\bga\tau}\\
m\bOm_{\bbe\tau}' & m\bOm_{\bga\tau}' & m\Omega_{\tau\tau}
\end{array}\right)=
\lim_{m\to\infty} {1\over m} \sum_{i=1}^m 
\left(\begin{array}{ccc}
 E[\psi_i^{\bbe}\psi_i^{\bbe'}] & E[\psi_i^{\bbe}\psi_i^{\bga'}] &  E[\psi_i^{\bbe}\psi_i^{\tau}]\\
 E[\psi_i^{\bga}\psi_i^{\bbe'}] &  E[\psi_i^{\bga}\psi_i^{\bga'}] &  E[\psi_i^{\bga}\psi_i^{\tau}]\\
 E[\psi_i^{\tau}\psi_i^{\bbe'}] &  E[\psi_i^{\tau}\psi_i^{\bga'}] &  E[\psi_i^{\tau}\psi_i^{\tau}]
\end{array}\right).
\end{align*}
It is noticed that $E[u_{1i}(y_{ij}-\x_{ij}'\bbe)]=0$ and $E[u_{2i}(y_{ij}-\x_{ij}'\bbe)]=0$ when $y_{ij}$ are normally distributed.
In such a case, it follows $\bOm_{\bbe\bga}=0$ and $\bOm_{\bbe\tau}=\0$, namely $\bbe$ and $\bphi=(\bga',\tau^2)'$ are asymptotically orthogonal.
However, since we do not assume the normality for observations $y_{ij}$'s, $\bbe$ and $\bphi$ are not necessarily orthogonal.

The asymptotic covariance matrix $m \bOm$ or $\bOm$ can be easily estimated from samples.
For example, $m\bOm_{\bbe\bbe}= \lim_{m\to\infty}m^{-1}\sum_{i=1}^m E[\psi_i^{\bbe}\psi_i^{\bbe'}]$ can be estimated by
$$
m \widehat{\bOm}_{\bbe\bbe}=\frac1{m}\sum_{i=1}^m\widehat{\psi_i^{\bbe}}\widehat{\psi_i^{\bbe'}},
$$
where $\widehat{\psi_i^{\bbe}}$ is obtained by replacing unknown parameters $\bth$ in $\psi_i^{\bbe}$ with estimates $\bthh$.
It is noted that the accuracy of estimation is given by
$$
\widehat{\bOm}_{\bbe\bbe}=\bOm_{\bbe\bbe}+o_p(m^{-1}), 
$$
from Theorem \ref{thm:asymp} and $\bOm=O(m^{-1})$.
The estimator $\widehat{\bOm}$ will be used to get the estimators of mean squared errors of predictors in Section \ref{sec:mse}.

We next provide the asymptotic properties of conditional covariance matrix given in the following corollary where the proof is given in the Appendix.

\begin{cor}\label{cor:cond}
Under Assumption {\rm (A)}, for $i=1,\ldots,m$, it follows that
\begin{equation}\label{cond}
E\left((\bthh-\bth)(\bthh-\bth)'\Big | \y_i\right)=\bOm+c(\y_i)o(m^{-1}),
\end{equation}
where $c(\y_i)$ is the fourth-order function of $\y_i$, so that $E|c(\y_i)|<\infty$ under Assumption (A).
\end{cor}

This property is used for estimation and evaluating the mean squared errors of EBLUP discussed in the subsequent section.
Moreover, in the evaluation of the mean squared errors of EBLUP and the derivation of its estimators, we need to obtain the conditional and unconditional asymptotic biases of the estimators $\bthh$.

Let $\b_{\bbe}^{(i)}(\y_i), \b_{\bga}^{(i)}(\y_i)$ and $\b_{\tau}^{(i)}(\y_i)$ be the second-order conditional asymptotic biases defined as
\begin{align*}
E[\bbeh-\bbe|\y_i]=&\b_{\bbe}^{(i)}(\y_i)+o_p(m^{-1}), \ \ E[\bgah-\bga|\y_i]=\b_{\bga}^{(i)}(\y_i)+o_p(m^{-1}), \\
&E[\tah^2-\tau^2|\y_i]=\b_{\tau}^{(i)}(\y_i)+o_p(m^{-1}).
\end{align*}
In the following theorem, we provide the analytical expressions of $\b_{\bbe}^{(i)}(\y_i), \b_{\bga}^{(i)}(\y_i)$ and $\b_{\tau}^{(i)}(\y_i)$.
Define $\b_{\bbe}$, $\b_{\bga}$ and $b_{\tau}$ by
\begin{align*}
\b_{\bbe}&=\left(\X'\bSi^{-1}\X\right)^{-1}\bigg\{\sum_{s=1}^q\sum_{k=1}^m\X_k'\bSi_k^{-1}\W_{i(s)}\bSi_k^{-1}\X_k\left(\bOm_{\beta^{\ast}\ga_s}-\bOm_{\beta\ga _s}\right)\\
&\ \ \ \ \ \ \ +\sum_{k=1}^m\X_k'\bSi_k^{-1}\J_{n_k}\bSi_k^{-1}\X_k(\bOm_{\bbe^{\ast}\tau}-\bOm_{\bbe\tau})\bigg\}
\end{align*}
\begin{equation}\label{bias}\begin{split}
\b_{\bga}&=\T_2(\bga)
\bigg[2\sum_{k=1}^m{\rm col}\left\{\tr\left(\E_k\Z_{kr}\E_k\X_k\left[\V_{\rm OLS}\X_k'-(\X'\X)^{-1}\X_k'\bSi_k\right]\right)\right\}_r\\
&\ \ \ \ \ \ \  -\sum_{k=1}^m\sum_{j=1}^{n_k}\z_{kj}\si_{kj(2)}^2(\z_{kj}-2n_k^{-1}\z_{kj}+n_k^{-1}\barz_k)'\bOm_{\bga\bga}\z_{kj}\bigg],
\end{split}\end{equation}
and
\begin{align*}
b_{\tau}&= - \frac1N\sum_{k=1}^m\sum_{j=1}^{n_k}\si_{kj(1)}^2\z_{jk}'\b_{\ga}-\frac{2}{N}\sum_{k=1}^m\tr\left\{(\X'\X)^{-1}\X_k'\bSi_k\X_k\right\}\\
&\ \ \ \ \ \ -\frac1{2N}\sum_{k=1}^m\sum_{j=1}^{n_k}\si_{kj(2)}^2\z_{kj}'\bOm_{\bga\bga}\z_{kj}+\frac1N\sum_{k=1}^m\tr\left(\X_k'\X_k\V_{\rm OLS}\right),
\end{align*}
where $\E_k=\I_{n_k}-n_k^{-1}\J_{n_k}$, $\V_{\rm OLS}=(\X'\X)^{-1}\X'\bSi\X(\X'\X)^{-1}$, $\Z_{kr}=\diag(z_{k1r},\ldots,z_{kn_kr})$ for $r$-th element $z_{kjr}$ of $\z_{kj}$, $\bOm_{\bbe^{\ast}a}$ for $a\in \{\tau,\ga_1,\ldots,\ga_q\}$ and $\W_{i(s)}$ are defined in the proof of Theorem \ref{thm:cond.bias}, and ${\rm col}\{a_r\}_r$ denotes a $q$-dimensional vector $(a_1,\ldots,a_q)'$.
It is noted that $\b_{\bbe},\b_{\bga},b_{\tau}$ are of order $O(m^{-1})$.
Now we provide the second-order approximation to the conditional asymptotic bias.

\begin{thm}\label{thm:cond.bias}
Under Assumption {\rm (A)}, we have
\begin{equation}\label{cond.bias}\begin{split}
\b_{\bbe}^{(i)}(\y_i)&=\left(\X'\bSi^{-1}\X\right)^{-1}\X_i'\bSi_i^{-1}(\y_i-\X_i\bbe)+\b_{\bbe}, \ \ \ \b_{\bga}^{(i)}(\y_i)=\T_2(\bga)\u_{2i}+\b_{\bga}\\
b_{\tau}^{(i)}(\y_i)&=m^{-1}u_{1i}-m^{-1}\T_1(\bga)'\T_2(\bga)\u_{2i}+ b_{\tau},
\end{split}\end{equation}
where $\b_{\bbe}^{(i)}(\y_i)$, $\b_{\bga}^{(i)}(\y_i)$ and $b_{\tau}^{(i)}(\y_i)$ are of order $O_p(m^{-1})$, and $u_{1i}$ and $u_{2i}$ are given in $(\ref{u1})$ and $(\ref{u2})$, respectively.
\end{thm}

From the above theorem, we immediately obtain the unconditional asymptotic bias of the estimators $\bthh$ by taking expectation with respect to $\y_i$ given in the following Corollary.

\begin{cor}\label{cor:bias}
Under Assumption {\rm (A)}, it holds that
$$
E[\bthh-\bth]=(\b_{\bbe}',\b_{\bga}',b_{\tau})'+o(m^{-1}),
$$
where $\b_{\bbe}$, $\b_{\bga}$ and $b_{\tau}$ are given in $(\ref{bias})$.
\end{cor}

\section{Prediction and Risk Evaluation}\label{sec:mse}

\subsection{Empirical predictor}
We now consider the prediction of 
$$
\mu_i=\c_i'\bbe+v_i,
$$
where $\c_i$ is a known (user specified) vector and $v_i$ is the random effect in model (\ref{model}).
The typical choice of $\c_i$ is $\c_i=\barx_i$ which corresponds to the prediction of mean of the $i$-th cluster.
A predictor $\mut(\y_i)$ of $\mu_i$ is evaluated in terms of the MSE $E[(\mut(\y_i)-\mu_i)^2]$. 
In the general forms of $\mut(\y_i)$, the minimizer (best predictor) of the MSE cannot be obtain without a distributional assumption for $v_i$ and $\ep_{ij}$.
Thus we focus on the class of linear and unbiased predictors, and the best linear unbiased predictor (BLUP) of $\mu_i$ in terms of the MSE is given by
$$
\mut_i=\c_i'\bbe+\1_{n_i}'\bSi_i^{-1}(\y_i-\X_i\bbe).
$$
This can be simplified as
$$
\mut_i=\c_i'\bbe+\sum_{j=1}^{n_i}\la_{ij}\left(y_{ij}-\x_{ij}'\bbe\right),
$$
where $\la_{ij}=\tau^2\si_{ij}^{-2}\eta_i^{-1}$ for $\eta_i=1+\tau^2\sum_{h=1}^{n_i}\si_{ih}^{-2}$.
In the case of homogeneous variances, namely $\si_{ij}^2=\delta^2$, it is confirmed that the BLP reduces to $\mut_i=\c_i'\bbe+\la_i\left(\bary_i-\barx_i'\bbe\right)$ with $\la_i=n_i\tau^2(\delta^2+n_i\tau^2)^{-1}$ as given in Hall and Maiti (2006).
The BLUP is not feasible since it depends on unknown parameters $\bbe$, $\bga$ and $\tau^2$. 
Plugging the estimators into $\mut_i$, we get the empirical best linear unbiased predictor (EBLUP) 
\begin{equation}\label{eblup}
\muh_i=\c_i'\bbeh+\sum_{j=1}^{n_i}\lah_{ij}\left(y_{ij}-\x_{ij}'\bbeh\right), \ \ \ \ \ \lah_{ij}=\tah^2\sih_{ij}^{-2}\etah_{i}^{-1}
\end{equation}
for $\etah_i^{-1}=1+\tah^2\sum_{h=1}^{n_i}\sih_{ih}^{-2}$.
In the subsequent section, we consider the mean squared errors (MSE) of EBLUP (\ref{eblup}) without any distributional assumptions for $v_i$ and $\ep_{ij}$.

\subsection{Second-order approximation to MSE}
To evaluate uncertainty of EBLUP given by (\ref{eblup}), we evaluate the MSE defined as ${\rm MSE}_i(\bphi)=E\left[(\muh_i-\mu_i)^2\right]$ for $\bphi=(\bga',\tau^2)'$.
The MSE is decomposed as
\begin{align*}
{\rm MSE}_i(\bphi)&=E\left[(\muh_i-\mut_i+\mut_i-\mu_i)^2\right]\\
&=E\left[(\mut_i-\mu_i)^2\right]+E\left[(\muh_i-\mut_i)^2\right]+2E\left[(\muh_i-\mut_i)(\mut_i-\mu_i)\right].
\end{align*}
From the expression of $\mut_i$, we have
$$
\mut_i-\mu_i=\left(\sum_{j=1}^{n_i}\la_{ij}-1\right)v_i+\sum_{j=1}^{n_i}\la_{ij}\ep_{ij},
$$
which leads to
\begin{equation}\label{R1}
R_{1i}(\bphi)\equiv E\left[(\mut_i-\mu_i)^2\right]=\left(\sum_{j=1}^{n_i}\la_{ij}-1\right)^2\tau^2+\sum_{j=1}^{n_i}\la_{ij}^2\si_{ij}^2=\tau^2\eta_i^{-1}.
\end{equation}

For the second term, however, we cannot obtain an exact expression, so that we derive the approximation up to $O(m^{-1})$.
Using the Taylor series expansion, we have
\begin{equation}\label{exp1}
\muh_i-\mut_i=\left(\frac{\partial\mut_i}{\partial\bth}\right)'(\bthh-\bth)+\frac12(\bthh-\bth)'\left(\frac{\partial^2\mut_i}{\partial\bth\partial\bth'}\Big|_{\bth=\bth^{\ast}}\right)(\bthh-\bth),
\end{equation}
where $\bth^{\ast}$ is on the line between $\bth$ and $\bthh$.
The straightforward calculation shows that
\begin{align}\label{mu.deriv}
\frac{\partial\mut_i}{\partial\bbe}=\c_i-\sum_{j=1}^{n_i}\la_{ij}\x_{ij}, \ \ \ \frac{\partial\mut_i}{\partial\bga}=\eta_i^{-2}\sum_{j=1}^{n_i}\si_{ij}^{-2}\bde_{ij}(y_{ij}-\x_{ij}'\bbe), \ \ \ \frac{\partial\mut_i}{\partial\tau^2}=\eta_{i}^{-2}\sum_{j=1}^{n_i}\si_{ij}^{-2}(y_{ij}-\x_{ij}'\bbe),
\end{align}
where
$$
\bde_{ij}=\tau^4\sum_{h=1}^{n_i}\siih^{-4}\si_{ih(1)}^2\z_{ih}-\tau^2\eta_i\siij^{-2}\si_{ij(1)}^2\z_{ij}.
$$
Then each element in $\partial^2\mut_i/\partial\bth\partial\bth'$ is a linear function of $\y_i$.
Hence under Assumption (A), using the similar arguments given in Lahiri and Rao (1995), we can show that
\begin{equation}\label{R2.proof}
E\left[(\muh_i-\mut_i)^2\right]=R_{2i}(\bphi)+o(m^{-1}),
\end{equation}
where the detailed proof is given in the Appendix, and 
\begin{equation}\begin{split}
R_{2i}(\bphi)=&\eta_i^{-4}\tau^2\left(\sum_{j=1}^{n_i}\siij^{-2}\bde_{ij}\right)'\bOm_{\ga\ga}\left(\sum_{j=1}^{n_i}\siij^{-2}\bde_{ij}\right)+\eta_i^{-4}\sum_{j=1}^{n_i}\siij^{-2}\bde_{ij}'\bOm_{\ga\ga}\bde_{ij}\\
&+2\eta_i^{-3}\sum_{j=1}^{n_i}\siij^{-2}\bde_{ij}'\bOm_{\ga\tau}+\eta_i^{-3}\sum_{j=1}^{n_i}\siij^{-2}\Omega_{\tau\tau}+\left(\c_i-\sum_{j=1}^{n_i}\la_{ij}\x_{ij}\right)'\bOm_{\bbe\bbe}\left(\c_i-\sum_{j=1}^{n_i}\la_{ij}\x_{ij}\right),
\label{R2}
\end{split}\end{equation}
which is of order $O(m^{-1})$.
All the evaluations of the residual terms appeared in this paper can be done by the similar manner, and detailed proofs will be omitted in what follows.

We next evaluate the cross term $E\left[(\muh_i-\mut_i)(\mut_i-\mu_i)\right]$. 
This term vanishes under the normality assumptions for $v_i$ and $\ep_{ij}$, but in general, it cannot be neglected.
As in the case of $R_{2i}$, we obtain an approximation of $E\left[(\muh_i-\mut_i)(\mut_i-\mu_i)\right]$ up to $O(m^{-1})$.
Noting that
$$
\mut_i-\mu_i=\left(\sum_{j=1}^{n_i}\la_{ij}-1\right)v_i+\sum_{j=1}^{n_i}\la_{ij}\ep_{ij}\equiv w_i,
$$
and using the expansion (\ref{exp1}), we obtain
\begin{align*}
E\left[(\muh_i-\mut_i)(\mut_i-\mu_i)\right]&=E\left[\left(\frac{\partial\mut_i}{\partial\bth}\right)'(\bthh-\bth)w_i\right]+\frac12E\left[(\bthh-\bth)'\left(\frac{\partial^2\mut_i}{\partial\bth\partial\bth'}\Big|_{\bth=\bth^{\ast}}\right)(\bthh-\bth)w_i\right].
\end{align*}
Using the expression of (\ref{mu.deriv}) and Corollary \ref{cor:cond}, the straightforward calculation (whose details are given in the Appendix) shows that
$$
R_{32i}(\bphi)\equiv \frac12E\left[(\bthh-\bth)'\left(\frac{\partial^2\mut_i}{\partial\bth\partial\bth'}\Big|_{\bth=\bth^{\ast}}\right)(\bthh-\bth)w_i\right]=o(m^{-1}),
$$
under Assumption (A). 
Moreover, from Theorem \ref{thm:cond.bias}, we obtain
$$
E\left[\left(\frac{\partial\mut_i}{\partial\bth}\right)'(\bthh-\bth)w_i\right]=R_{31i}(\bphi,\bka)+o(m^{-1}),
$$
for
\begin{equation}\label{R31}\begin{split}
R_{31i}(\bphi,\bka)&=\eta_i^{-2}\sum_{j=1}^{n_i}\siij^{-2}\bde_{ij}'\left(\sum_{k=1}^{m}\sum_{h=1}^{n_k}\si_{kh(1)}^2\z_{kh}\z_{kh}'\right)^{-1}\M_{2ij}(\bphi,\bka)\\
&\ \ \ \ +m^{-1}\eta_i^{-2}\sum_{j=1}^{n_i}\siij^{-2}\bigg\{M_{1ij}(\bphi,\bka)-\T_1(\bga)'\T_2(\bga)\M_{2ij}(\bphi,\bka)\bigg\},
\end{split}\end{equation}
where
\begin{equation*}\begin{split}
&M_{1ij}(\bphi,\bka)=mN^{-1}\tau^2\eta_i^{-1}\Big\{n_i\tau^2(3-\ka_v)+\siij^2(\ka_{\ep}-3)\Big\}\\
&\M_{2ij}(\bphi,\bka)=mN^{-1}\tau^2\eta_i^{-1}n_i^{-2}(n_i-1)^2(\ka_{\ep}-3)\siij^2\z_{ij},
\end{split}\end{equation*}
and $\ka_{v}$, $\ka_{\ep }$ are defined as $E(v_i^4)=\kappa_{v}\tau^4$ and $E(\ep_{ij}^4)=\kappa_{\ep }\siij^4$, respectively, and $\bka=(\ka_{v},\ka_{\ep})'$.
The derivation of the expression of $R_{31i}(\bphi,\bka)$ is also given in the Appendix.
From the expression (\ref{R31}), it holds that $R_{31i}(\bphi,\bka)=O(m^{-1})$.

Under the normality assumption of $v_i$ and $\ep_{ij}$, we immediately obtain $M_{1ij}=0$ and $\M_{2ij}=\0$ since $\bka=(3,3)'$.
This leads to $R_{31}=0$, which means that the cross term does not appear in the second-order approximated MSE, that is our result is consistent to the well-known result.

Now, we summarize the result for the second-order approximation of the MSE.

\begin{thm}\label{thm:mse}
Under Assumption {\rm (A)}, the second-order approximation of the MSE is given by
$$
{\rm MSE}_i(\bphi)=R_{1i}(\bphi)+R_{2i}(\bphi)+2R_{31i}(\bphi,\bka)+o(m^{-1}),
$$
where $R_{1i}(\bphi)$, $R_{2i}(\bphi)$ and $R_{31i}(\bphi,\bka)$ are given in $(\ref{R1})$, $(\ref{R2})$ and $(\ref{R31})$, respectively, and $R_{1i}(\bphi)=O(1)$, $R_{2i}(\bphi)=O(m^{-1})$ and $R_{31i}(\bphi,\bka)=O(m^{-1})$.
\end{thm}

The approximated MSE given in Theorem \ref{thm:mse} depends on unknown parameters.
Thus, in the subsequent section, we derive the second-order unbiased estimator of the MSE by the analytical and the matching bootstrap methods.

\subsection{Analytical estimator of the MSE}

We first derive the analytical second-order unbiased estimator of the MSE.
From Theorem \ref{thm:mse}, $R_{2i}(\bphi)$ is $O(m^{-1})$, so that it can be estimated by the plug-in estimator $R_{2i}(\bphih)$ with second-order accuracy, namely $E[R_{2i}(\bphih)]=R_{2i}(\bphi)+o(m^{-1})$.
For $R_{31i}(\bphi,\bka)$ with order $O(m^{-1})$, if a consistent estimator $\bkah$ is available for $\bka$, this term can be estimated by the plug-in estimator with second-order unbiasedness.
To this end, we construct a consistent estimator of $\bka$ using the expression of fourth moment of observations.
The straightforward calculation shows that
\begin{align*}
E&\left[\sum_{j=1}^{n_i}\left\{y_{ij}-\bary_i-(\x_{ij}-\barx_i)'\bbe\right\}^4\right]\\
&\ \ \ =\kappa_{\ep}n_i^{-4}(n_i-1)(n_i-2)(n_i^2-n_i-1)\left(\sum_{j=1}^{n_i}\siij^4\right)+3n_i^{-3}(2n_i-3)\left\{\left(\sum_{j=1}^{n_i}\siij^2\right)^2-\sum_{j=1}^{n_i}\siij^4\right\},
\end{align*}
whereby we can estimate $\ka_{\ep}$ by
\begin{equation}\label{ka.ep}\begin{split}
\kah_{\ep }&=\frac1{N^{\ast}}\sum_{i=1}^m\left[\sum_{j=1}^{n_i}\left\{y_{ij}-\bary_i-(\x_{ij}-\barx_i)'\bbeh\right\}^4-3n_i^{-3}(2n_i-3)\left\{\left(\sum_{j=1}^{n_i}\siij^2\right)^2-\sum_{j=1}^{n_i}\siij^4\right\}\right],
\end{split}\end{equation}
where $N^{\ast}=n_i^{-4}(n_i-1)(n_i-2)(n_i^2-n_i-1)\sum_{j=1}^{n_i}\siij^4$ and $\bbeh$ is the feasible GLS estimator of $\bbe$ given in Section 2.
For $\ka_{v}$, it is observed that
\begin{align*}
E\left[\left(y_{ij}-\x_{ij}'\bbe\right)^4\right]=\tau^4\ka_{v}+6\tau^2\siij^2+\ka_{\ep}\siij^4,
\end{align*}
which leads to the estimator of $\ka_{v}$ given by
\begin{equation}\label{ka.v}\begin{split}
\kah_{v}&=\frac1{N\tah^4}\sum_{i=1}^m\sum_{j=1}^{n_i}\left\{\left(y_{ij}-\x_{ij}'\Obbeh\right)^4-6\tah^2\sih_{ij}^2-\kah_{\ep}\sih_{ij}^4\right\}.
\end{split}\end{equation}
From Theorem \ref{thm:asymp}, it immediately follows that the estimators given in (\ref{ka.ep}) and (\ref{ka.v}) are consistent.
Using these estimators, we can estimate $R_{31i}$ by $R_{31i}(\bphih,\bkah)$ with second-order accuracy.

Finally, we consider the second-order unbiased estimation of $R_{1i}$.
The situation is different than before since $R_{1i}=O(1)$, which means that the plug-in estimator $R_{1i}(\bphih)$ has the second-order bias with $O(m^{-1})$. 
Thus we need to obtain the second-order bias of $R_{1i}(\bphih)$ and correct them.
By the Taylor series expansion, we have
$$
R_{1i}(\bphih)=R_{1i}(\bphi)+\left(\frac{\partial R_{1i}(\bphi)}{\partial\bphi'}\right)(\bphih-\bphi)+\frac12(\bphi-\bphi)'\left(\frac{\partial^2 R_{1i}(\bphi)}{\partial\bphi\partial\bphi'}\right)(\bphih-\bphi)+o_p(\|\bphih-\bphi\|^2).
$$
Then, the second-order bias of $R_{1i}(\bphih)$ is expressed as
\begin{align*}
E[R_{1i}(\bphih)]&-R_{1i}(\bphi)\\
&=\left(\frac{\partial R_{1i}(\bphi)}{\partial\bphi'}\right)E[\bphih-\bphi]+\frac12\tr\left\{\left(\frac{\partial^2 R_{1i}(\bphi)}{\partial\bphi\partial\bphi'}\right)E\left[(\bphih-\bphi)(\bphih-\bphi)'\right]\right\}+o(m^{-1})\\
&=\left(\frac{\partial R_{1i}(\bphi)}{\partial\bphi'}\right)\b_{\bphi}+\frac12\tr\left\{\left(\frac{\partial^2 R_{1i}(\bphi)}{\partial\bphi\partial\bphi'}\right)\bOm_{\bphi}\right\}+o(m^{-1}),
\end{align*}
where $\bOm_{\bphi}$ is the sub-matrix of $\bOm$ with respect to $\bphi$, and $\b_{\phi}$ is the second-order bias of $\bphih$ given in Corollary \ref{cor:bias}.
The straightforward calculation shows that
\begin{align*}
\frac{\partial R_{1i}(\bphi)}{\partial\tau^2}&=\eta_i^{-2}, \ \ \ \ \frac{\partial R_{1i}(\bphi)}{\partial\bga}=-\tau^2\eta_i^{-2}\bta_{i(1)}, \ \ \ \ \frac{\partial^2 R_{1i}(\bphi)}{\partial\tau^2\partial\tau^2}=2\tau^{-2}(\eta_i^{-3}-\eta_i^{-2}),\\
\frac{\partial^2 R_{1i}(\bphi)}{\partial\bga\partial\tau^2}&=-2\eta_i^{-3}\bta_{i(1)}, \ \ \ \ 
\frac{\partial^2 R_{1i}(\bphi)}{\partial\bga\partial\bga'}=\tau^2\eta_i^{-3}(2\bta_{i(1)}\bta_{i(1)}'-\eta_i\bta_{i(2)}),
\end{align*}
where
$$
\bta_{i(1)}\equiv \frac{\partial\eta_i}{\partial\bga}=-\tau^2\sum_{j=1}^{n_i}\siij^{-4}\si_{ij(1)}^2\z_{ij}, \ \ \ \bta_{i(2)}\equiv \frac{\partial^2\eta_i}{\partial\bga\partial\bga'}=\tau^2\sum_{j=1}^{n_i}\left(2\siij^{-2}\si_{ij(1)}^4-\si_{ij(2)}^2\right)\siij^{-4}\z_{ij}\z_{ij}'.
$$
Therefore, we obtain the expression of the second-order bias given by
\begin{equation}\label{R.bais}
\begin{split}
B_{i}(\bphi)=&-\tau^2\eta_i^{-2}\bta_{i(1)}'\b_{\ga}+\eta_i^{-2}b_{\tau}-2\eta_i^{-3}\bta_{i(1)}'\bOm_{\ga\tau}+\tau^{-2}(\eta_i^{-3}-\eta_i^{-2})\Omega_{\tau\tau}\\
&+\tau^2\eta_i^{-3}\left\{\bta_{i(1)}'\bOm_{\ga\ga}\bta_{i(1)}-\frac12\eta_i\tr\left(\bta_{i(2)}\bOm_{\ga\ga}\right)\right\},
\end{split}\end{equation}
with $B_{i}(\bphi)=O(m^{-1})$.
Noting that $B_{i}(\bphi)$ can be estimated by $B_i(\bphih)$ with $E[B_i(\bphih)]=B_{i}(\bphi)+o(m^{-1})$ from Theorem \ref{thm:asymp}, we propose the bias corrected estimator of $R_{1i}$ given by 
\begin{equation*}\label{R1.est}
\widehat{R_{1i}}(\bphih)^{bc}=R_{1i}(\bphih)-B_i(\bphih),
\end{equation*}
which is second-order unbiased estimator of $R_{1i}$, namely 
$$
E[\widehat{R_{1i}}(\bphih)^{bc}]=R_{1i}(\bphi)+o(m^{-1}).
$$
Now, we summarize the result for the second-order unbiased estimator of MSE in the following theorem.

\begin{thm}\label{thm:mseest}
Under Assumption {\rm (A)}, the second-order unbiased estimator of ${\rm MSE}_i$ is given by
$$
\widehat{\rm MSE}_i=\widehat{R_{1i}}(\bphih)^{bc}+R_{2i}(\bphih)+2R_{31i}(\bphih,\bkah),
$$
that is, $E\left[\widehat{\rm MSE}_i\right]={\rm MSE}_i+o(m^{-1})$.
\end{thm}

It is remarked that the proposed estimator of MSE does not require any resampling methods such as bootstrap.
This means that the analytical estimator can be easily implemented and has less computational burden compared to bootstrap.
Moreover, we do not assume normality of $v_i$ and $\ep_{ij}$ in the derivation of the MSE estimator as in Lahiri and Rao (1995).
Thus the proposed MSE estimator is expected to have a robustness property, which will be investigated in the simulation studies.

\section{Simulation and Empirical Studies}\label{sec:sim}

\subsection{Model based simulation}\label{sec:comp}
We first compare the performances of EBLUP obtained from the proposed HNER with variance functions (HNERVF) with several existing models in terms of simulated mean squared errors (MSE).
We consider the conventional nested error regression (NER) model, heteroscedastic NER model given by Jiang and Nguyen (2012) referred as JN, and the heteroscedastic NER with random dispersions (HNERRD) proposed in Kubokawa, et al. (2016).
In applying the NER model, we use the unbiased estimator for variance components given in Prasad and Rao (1990) to calculate EBLUP.
Further, we also consider the following log-link gamma mixed (GM) models as the competitor from the generalized linear mixed models, which also allows heteroscedasticity for the variances as the quadratic function of means.
We used {\tt glmer} function in {\tt lme4} package in `R' to apply the GM model.

In this simulation study, we set $m=20$ and $n_i=8$ in all cases, and we compute the simulated MSE in 10 scenarios denoted by S1$,\ldots,$S10.
The simulated MSE for some area-specific parameter $\mu_i$ is define as 
\begin{equation}\label{sim-MSE}
{\rm MSE}_i=\frac1R\sum_{r=1}^R(\muh_i^{(r)}-\mu_i^{(r)})^2,
\end{equation}
where $R=5000$ is the number of simulation runs, $\muh_i^{(r)}$ is the predicted value from some models and $\mu_i^{(r)}$ is the true values in the $r$-th iteration.
In all scenarios, we generate covariates $x_{ij}$'s from the uniform distribution on $(0,1)$, which are fixed in simulation runs.
From S1 to S3, we consider the heteroscedastic model with area-level heteroscedastic variances given by
\begin{align*}
{\rm S1}\sim {\rm S3}: \  \ y_{ij}=\be_0+\be_1x_{ij}+v_i+\ep_{ij}, \  \ v_i\sim (0,\tau^2), \ \ \ep_{ij}\sim (0,\si_i^2), \ \ \ \mu_i=\be_0+v_i,
\end{align*}
where $\si_i^2=\exp(0.8-z_i)$ and $(\beta_0,\beta_1,\tau)=(1,0.5,1.2)$.
We generate $z_i$'s from uniform distribution on $(-1,1)$, which are fixed in simulation runs.
The scenarios S1, S2 and S3 correspond to the cases where the distributions of both $v_i$ and $\ep_{ij}$ are normal, $t$ with 6 degrees of freedom, and chi-squared with 5 degrees of freedom, respectively, noting that both $t$-distribution and chi-squared distribution are scaled and located to meet the specified means and variances.
For S4, we consider the homoscedastic model given by
$$
{\rm S4}: \ \ \ y_{ij}=\be_0+\be_1x_{ij}+v_i+\ep_{ij}, \  \ v_i\sim N(0,\tau^2), \ \ \ep_{ij}\sim N(0,\si^2), \ \ \ \mu_i=\be_0+v_i,
$$ 
with $(\beta_0,\beta_1,\tau,\si)=(1,0.5,1.2,1.5)$.
In S5 and S6, we use the heteroscedastic model with unit-level heteroscedastic variances given by
$$
{\rm S5,S6}: \  \ y_{ij}=\be_0+\be_1x_{ij}+v_i+\ep_{ij}, \  \ v_i\sim N(0,\tau^2), \ \ \ep_{ij}\sim N(0,\si_{ij}^2), \ \ \ \mu_i=\be_0+v_i,
$$
where $\si_{ij}^2=\exp(0.8-z_{ij})$ in S5 and $\si_{ij}^2\sim \Gamma(5,5/\exp(0.8-z_{ij}))$ in S6. 
For S7 and S8, we consider the mixed model of the form
$$
{\rm S7,S8}: \ \ y_{ij}=\exp(\be_0+\be_1x_{ij}+v_i)\ep_{ij}, \ \  \ \mu_i=\exp(\be_0+v_i),
$$
where $v_i\sim N(0,\tau^2)$, $\ep_{ij}\sim \Gamma(3,3)$ and $(\beta_0,\beta_1,\tau)=(0.5,1,0.3)$ in S7, and $v_i\sim t_6(0,\tau^2)$, $\ep_{ij}\sim SLN(1,\si^2)$, and $(\beta_0,\beta_1,\tau,\si)=(1.2,0.6,0.4,0.4)$ in S8, noting that $t_6(a,b)$ denotes the $t$-distribution with $6$ degrees of freedom with mean $a$ and variance $b$ and $SLN(a,b)$ denotes the scaled log-normal distribution with mean $a$ and variance $b$.
Hence, $S7$ corresponds to the gamma mixed model with log-link function and $S8$ corresponds to its misspecified version.
Finally, S9 to S10 are the mixed models defined as 
$$
{\rm S9}: \ \ y_{ij}=(\be_0+\be_1x_{ij}+v_i)^2\ep_{ij}, \ \ \ v_i\sim N(0,\tau^2), \ \ \ \ep_{ij}\sim SLN(1,\si^2), \ \ \ \mu_i=(\be_0+v_i)^2
$$
with $(\beta_0,\beta_1,\tau,\si)=(1,0.6,1.5,0.5)$, and 
$$
{\rm S10}: \ \ y_{ij}=\{\exp(\be_0+\be_1x_{ij})+v_i\}\ep_{ij}, \ \ \ v_i\sim N(0,\tau^2), \ \ \ \ep_{ij}\sim SLN(1,\si^2), \ \ \ \mu_i=\exp(\be_0)+v_i,
$$
with $(\beta_0,\beta_1,\tau,\si)=(1,0.3,1.2,0.5)$.
It is noted that both S9 and S10 are also heteroscedastic model in the sense that $\Var(y_{ij})$ depends on $x_{ij}$.

Under the 10 scenarios described above, we compute the simulated MSE values of predictors from five methods (HNERVF, HNERRD, NER, JN and GM) in each area.
Since we can apply GM only to the data with positive $y_{ij}$'s, the MSE values of GM model are calculated from S7 to S10. 
In Table \ref{tab:comp}, we show the mean, max and min values of MSE over all areas for each model and scenario.
From S1 to S3, it is observed that HNERVF performs better than the other models, and NER model performs worst since the true model is heteroscedastic.
In S4, NER model performs best among four models since NER model is the true model and other HNER models are overfitted.
It is also interesting to point out that the inefficiency of the prediction of JN is more serious than that of HNERVF and HNERRD.
As in S5 and S6, the heteroscedastic variances are unit-level, the amount of improvement of HNERVF over other models gets greater.
The scenario S7 corresponds to GM model, so that it is reasonable that MSE of GM is smallest among five models.
The scenario S8 is not GM model but it is still close to GM model, in which GM model works well compared to the other models.
However, once GM is seriously misspecified as in S9 and S10, GM does not work very much because of its somewhat strong parametric assumption.
From S8 to S10, all models are misspecified, but HNERVF model works well compared to other models.
Therefore, it is natural that HNERVF performs best when HNERVF is the true model, but even in case that HNERVF is misspecified, HNERVF also works reasonably well owing to its flexible structure of the model.

\begin{table}[htbp]
\caption{Simulated Values of MSE for Various Scenarios and Models
\label{tab:comp}
}

\medskip
\begin{center}
\begin{tabular}{ccccccccccccccccc}
\toprule
  & model& S1& S2 &S3 & S4 & S5 & S6 & S7 &S8  & S9 & S10 \\
\midrule  
&HNERVF & 0.368 & 0.370 & 0.371 & 0.311 &0.280 &0.293 & 0.269 & 0.619& 0.198 & 0.376\\
&HNERRD & 0.383 & 0.383 & 0.387 & 0.310 & 0.341 & 0.379 & 0.285 & 0.641& 0.259 & 0.369\\
mean&NER & 0.398 & 0.405 & 0.410 & 0.307 & 0.342 & 0.384 & 0.375 & 0.726& 0.220 & 0.384\\
&JN & 0.386 & 0.392 & 0.396 & 0.324 & 0.357 & 0.392 & 0.292 & 0.684& 0.318 & 0.385\\
&GM & --- & ---& --- & ---& --- & ---& 0.130 & 0.451 & 0.231 & 0.396\\
\midrule  
&HNERVF & 0.598 & 0.633 & 0.569 & 0.340 & 0.354 & 0.469 & 0.342 & 1.511& 0.299 & 0.435\\
&HNERRD & 0.630 & 0.634 & 0.603 & 0.342 & 0.424 & 0.523 & 0.405 & 1.603& 0.415 & 0.419\\
max&NER  & 0.642 & 0.639 & 0.596 & 0.339 &  0.423 & 0.526 & 0.518 & 1.992& 0.336 & 0.439\\
&JN & 0.634 & 0.643 & 0.618 & 0.372 & 0.445 & 0.545 & 0.426 & 1.834& 0.532 & 0.441\\
&GM& --- & ---& --- & ---& --- & ---& 0.149 & 0.970&0.372 & 0.473\\
\midrule  
&HNERVF & 0.138 &0.145 & 0.150 & 0.272 & 0.202 & 0.196 & 0.205 & 0.398& 0.142 & 0.297\\
&HNERRD & 0.156 & 0.157 & 0.166 & 0.272 & 0.254 & 0.255 & 0.219 & 0.408& 0.142 & 0.302\\
min&NER & 0.173 & 0.177 & 0.202 & 0.269 & 0.256 & 0.256 & 0.286 & 0.442& 0.152 & 0.305\\
&JN & 0.157 & 0.160 & 0.166 & 0.288 & 0.273 & 0.256 & 0.220 & 0.414& 0.168 & 0.314\\
&GM& --- & ---& --- & ---& --- & ---& 0.104 & 0.335& 0.168 & 0.309\\
\bottomrule
\end{tabular}
\end{center}
\end{table}

\subsection{Finite sample performances of the MSE estimator}
We next investigate the finite sample performances of the MSE estimators given in Theorem \ref{thm:mseest}.
To this end, we consider the data generating process given by 
$$
y_{ij}=\be_{0}+\be_{1}x_{ij}+v_{i}+\ep_{ij}, \ \ \ 
v_{i}\sim (0,\tau^{2}), \ \ \ \ep_{ij}\sim (0,\exp(\ga_{0}+\ga_{1}z_{ij})) 
$$
with $\beta_0=1,\beta_1=0.8$, $\tau=1.2$, $\ga_0=1$ and $\ga_1=-0.4$.
Moreover, we equally divided $m=20$ areas into 5 groups ($G=1,\ldots,5$), so that each group has 4 areas and the areas in the same group has the same sample size $n_G=G+3$.
Following Hall and Maiti (2006), we consider five patterns of distributions of $v_i$ and $\ep_{ij}$, that is , M1: $v_i$ and $\ep_{ij}$ are both normally distributed, M2: $v_i$ and $\ep_{ij}$ are both scaled $t$-distribution with degrees of freedom $6$, M3: $v_i$ and $\ep_{ij}$ are both scaled and located $\chi_5$ distribution, M4: $v_i$ are $\ep_{ij}$ are scaled and located $\chi_5$ and $-\chi_5$ distribution, respectively, and M5: $v_i$ are $\ep_{ij}$ are both logistic distribution.
The simulated values of the MSE are obtained from (\ref{sim-MSE}) based on $R=10,000$ simulation runs.
Then, based on $R=5,000$ simulation runs, we calculate the relative bias (RB) and coefficient of variation (CV) of MSE estimators given by
$$
{\rm RB}_i=\frac1R\sum_{r=1}^{R}\frac{\widehat{\rm MSE}_i^{(r)}-{\rm MSE}_i}{{\rm MSE}_{i}},\ \ \ \ {\rm CV}_{i}^2=\frac1R\sum_{r=1}^{R}\left(\frac{\widehat{\rm MSE}_i^{(r)}-{\rm MSE}_i}{{\rm MSE}_{i}}\right)^{2}
$$
where $\widehat{\rm MSE}_i^{(r)}$ is the MSE estimator in the $r$-th iteration.
In Table \ref{mse-sim}, we report mean and median values of ${\rm RB}_i$ and ${\rm CV}_i$ in each group.
For comparison, results for the naive MSE estimator, without any bias correction, are reported in Table \ref{mse-sim} as RBN.
The naive MSE estimator is the plug-in estimator of the asymptotic MSE (\ref{R1}), namely it is obtained by replacing $\tau^2$ and $\bga$ in formula (\ref{R1}) by $\tah^2$ and $\bgah$, respectively.
In Table \ref{mse-sim}, the relative bias is small, less than 10\% in many cases.
When the underlying distributions leave from normality, the MSE estimator still provides small relative bias although it has higher coefficient of variation.
The naive MSE estimator is more biased than the analytical MSE estimator in all groups and models, so that the bias correction in MSE estimator is successful.

\begin{table}[htbp]
\caption{The Mean Values of Percentage Relative Bias (RB) and Coefficient of Variation (CV) of MSE Estimator and Relative Bias of Naive MSE Estimator (RBN) in Each Group. }
\label{mse-sim}
\medskip
\begin{center}
\scalebox{1}{
\begin{tabular}{ccccccccccccccccc}
\toprule
Group&& Measure && M1 & M2 & M3 & M4 & M5\\
\midrule
 && RB  && -8.72 & -12.50 & -10.86 & -11.51 & -11.81\\
 $G_1$ && CV &&17.48 & 23.60 & 23.47 & 23.40 & 21.24 \\
 && RBN && -12.67 & -13.74 & -13.10 & -13.57 & -13.39\\
 \midrule
 && RB  && -7.61 & -9.72 & -10.58 & -10.57 & -7.27\\
 $G_2$ && CV &&  17.52 & 23.24 & 22.70 & 23.03 & 20.31 \\
 && RBN && -10.16 & -12.66 & -11.48 & -11.33 & -10.54\\
 \midrule
 && RB  && -7.89 & -8.39 & -7.65 & -8.92 & -6.34\\
 $G_3$ && CV && 19.85 & 26.05 & 24.66 & 25.37 & 22.94 \\
 && RBN && -9.31 & -9.43 & -8.70 & -9.86 & -7.58 \\
 \midrule
 && RB  && -6.52 & -4.74 & -4.96 & -5.65 & -4.27\\
 $G_4$ && CV && 22.02 & 28.37 & 26.93 & 27.68 & 24.98 \\
 && RBN && -10.83 & -7.68 & -7.98 & -6.52 & -6.42\\
\bottomrule
\end{tabular}
}
\end{center}
\end{table}

\subsection{Real data application}\label{sec:app}
We now apply the HNERVF model together with HNERRD, NER, JN and GM models considered in the simulation study in Section \ref{sec:comp} to the data which originates from the posted land price (PLP) data along the Keikyu train line in 2001.
This train line connects the suburbs in the Kanagawa prefecture to the Tokyo metropolitan area.
Those who live in the suburbs in the Kanagawa prefecture take this line to work or study in Tokyo everyday, so that it is expected that the land price depends on the distance from Tokyo. 
The PLP data are available for 52 stations on the Keikyu train line, and we consider each station as a small area, namely, $m=52$.
For the $i$-th station, data of $n_i$ land spots are available, where $n_i$ varies around 4 and some areas have only one observation.

\begin{figure}[!thb]
\centering
\includegraphics[width=8.2cm,clip]{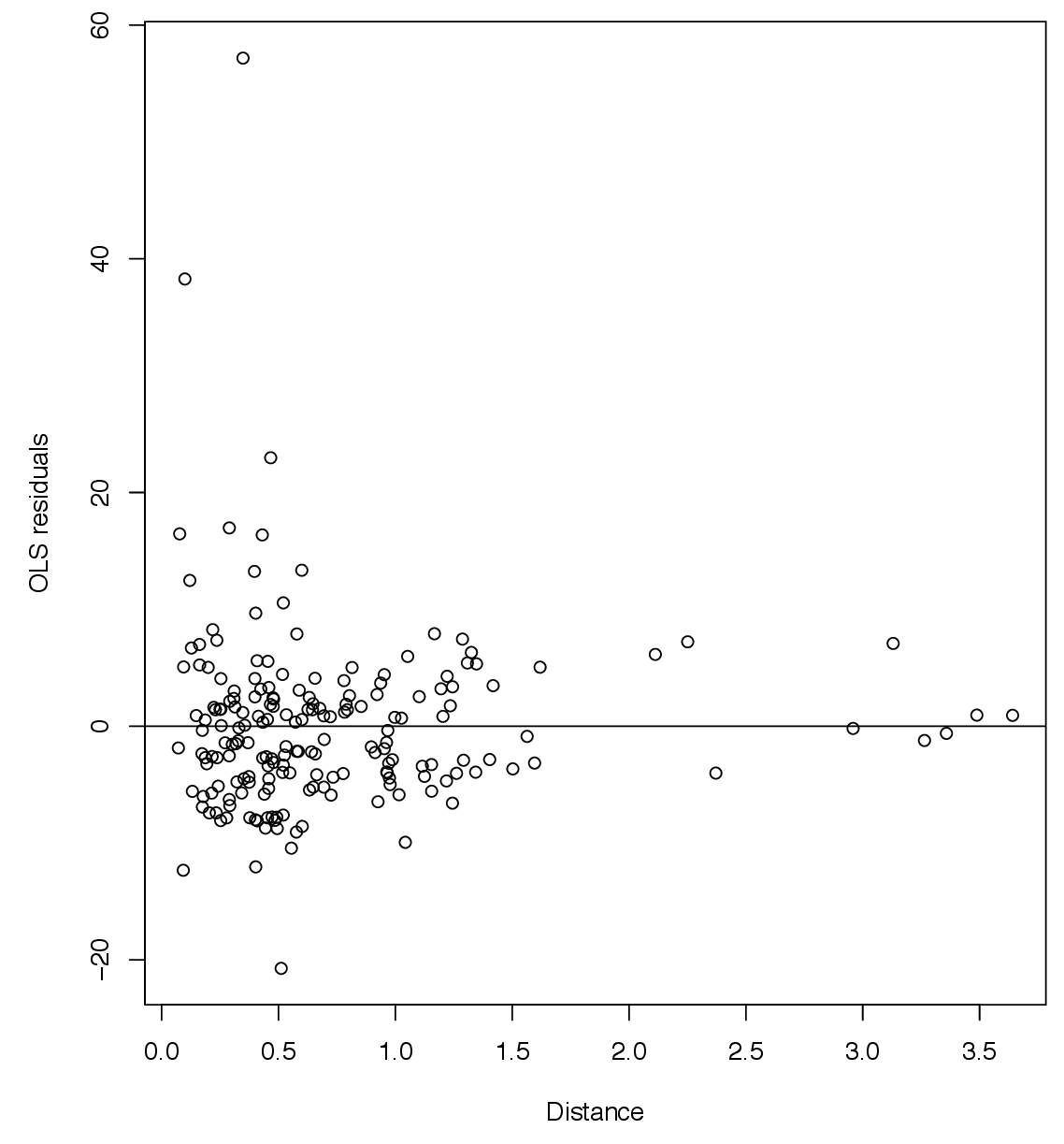}
\includegraphics[width=8.2cm,clip]{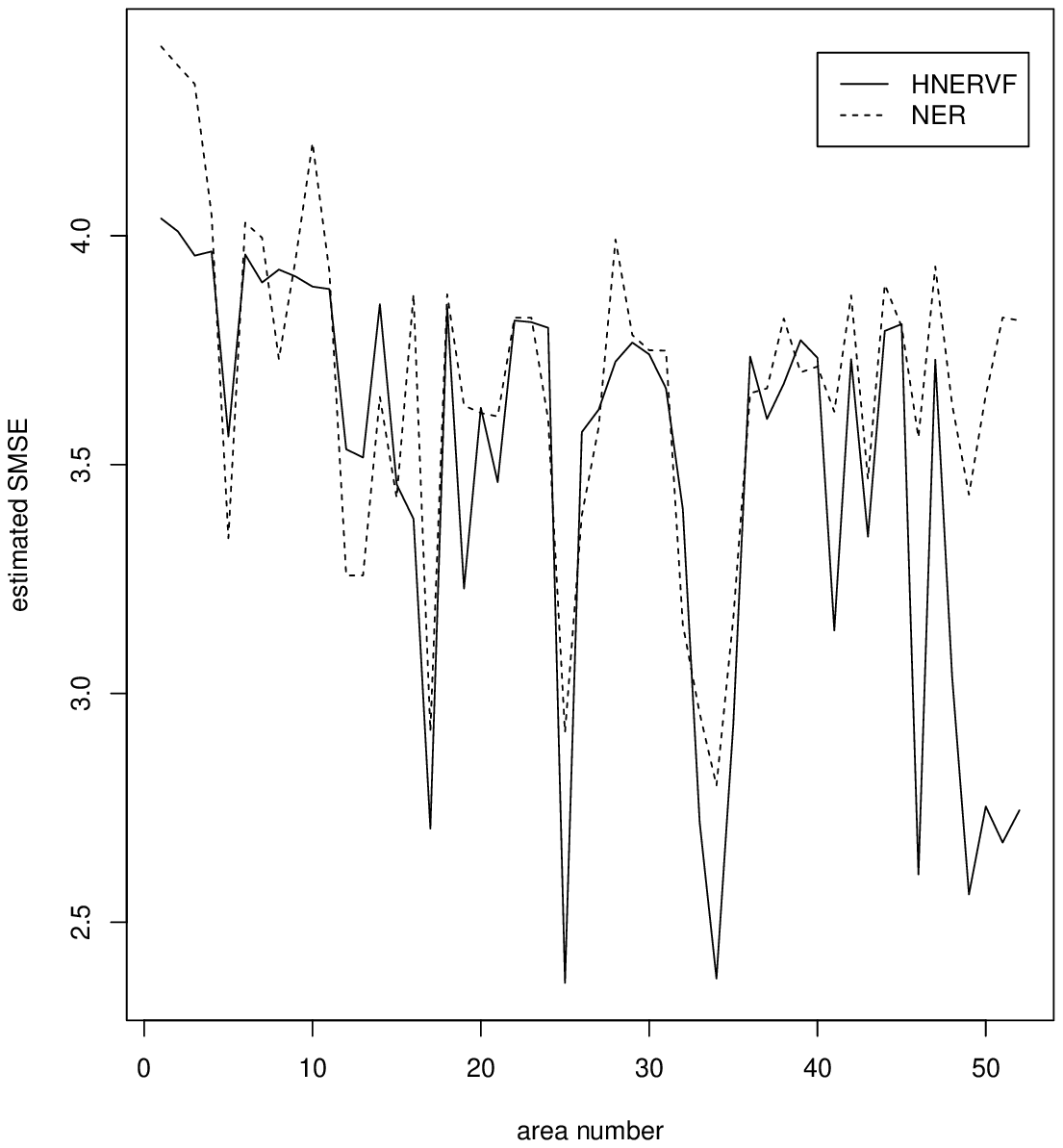}
\caption{Plot of OLS Residuals Against Distance $D_{ij}$ (Left) and Estimated root of MSE (RMSE) in HNERVF and NER models (Right). }
\label{fig:PLP}
\end{figure}

For $j=1, \ldots, n_i$, $y_{ij}$ denotes the scaled value of the PLP (Yen/10,000) for the unit meter squares of the $j$-th spot, $T_{i}$ is the time to take from the nearby station $i$ to the Tokyo station around 8:30 in the morning, $D_{ij}$ is the value of geographical distance from the spot $j$ to the station $i$ and $FAR_{ij}$ denotes the floor-area ratio, or ratio of building volume to lot area of the spot $j$. 
The three covariates $FAR_{ij}$, $T_i$ and $D_{ij}$ are also scaled by 100,10 and 1000, respectively.
This data set is treated in Kubokawa, et al. (2016), where they pointed out that the heteroscedasticity seem to be appropriate from boxplots of some areas and Bartlett test for testing homoscedastic variance.
They used the PLP data with log-transformed observations, namely $\log y_{ij}$, but we use $y_{ij}$ in this study since the results are easier to interpret than the results from $\log y_{ij}$.  
In the left panel of Figure \ref{fig:PLP}, we show the plot of the pairs $(D_{ij},e_{ij})$, where $e_{ij}$ is OLS residuals defined as 
$$
e_{ij}=y_{ij}-(\beh_{0,OLS} + FAR_{ij}\beh_{1,OLS} + T_{i}\beh_{2,OLS} + D_{ij}\beh_{3,OLS}).
$$
The figure indicates that the residuals are more variable for small $D_{ij}$ than for large $D_{ij}$, and the variances are exponentially decreasing with respect to $D_{ij}$.
Thus we apply the HNERVF model with the exponential variance function given by
\begin{equation}
y_{ij} = \be_0 + FAR_{ij}\be_1 + T_{i}\be_2 + D_{ij}\be_3  + v_i + \ep_{ij},
\label{PLP-model}
\end{equation}
where $v_{i}\sim (0,\tau^{2})$ and $\ep_{ij}\sim (0,\exp(\ga_{0}+\ga_{1}D_{ij}))$.
To compare the results, we also apply HNERRD, NER, JN and GM models to the PLP data with the same covariates.
In applying NER model, we regard it as the submodel of HNERVF by putting $\ga_1=0$ and use the same estimating method with HNERVF. 
The estimated regression coefficients from five models are given in the Table \ref{tab:PLPest}.
We first note that the conditional expectation of the GM model is $\exp(\be_0 + FAR_{ij}\be_1 + T_{i}\be_2 + D_{ij}\be_3  + v_i )$, while that of other models has the liner form $\be_0 + FAR_{ij}\be_1 + T_{i}\be_2 + D_{ij}\be_3  + v_i $.
Hence the scale of the estimated coefficients of GM are different from those of other models.
However, the signs of estimated coefficients are the same over all models.
The resulting signs are intuitively natural since the PLP is expected to be decreasing as the distance between the spot and the nearest station gets large or the nearest station gets distant from Tokyo station.
Moreover, in HNERVF model, the estimated value of $\ga_1$ is $\gah_1=-1.82$, which is consistent to the observation from the left panel of Figure \ref{fig:PLP}.
Using the result of Theorem \ref{thm:asymp}, the asymptotic standard error of $\gah_1$ is $0.492$, so that $\ga_1$ seems significant.

\begin{figure}[!thb]
\centering
\includegraphics[width=13cm,clip]{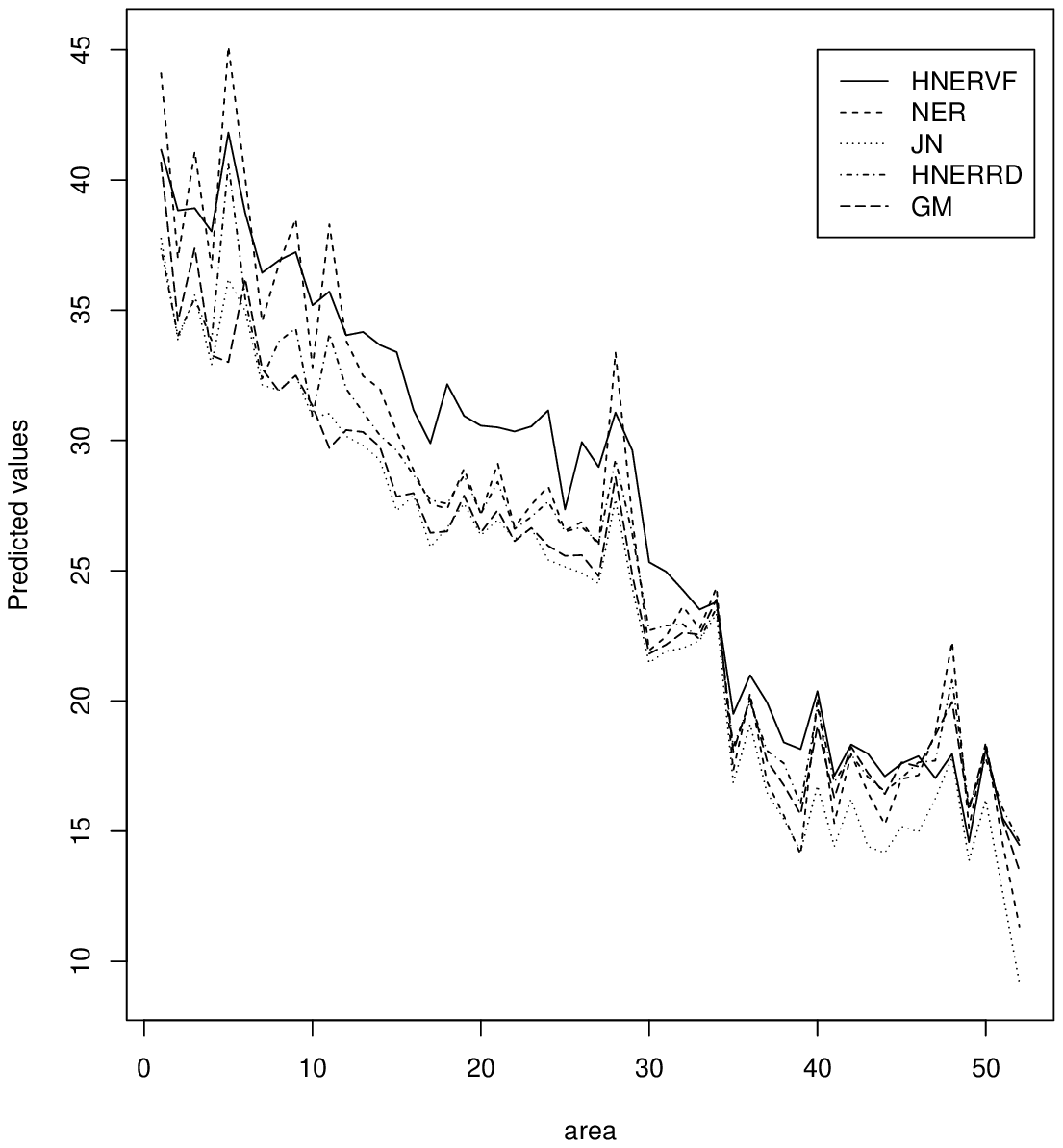}
\caption{Predicted Values of $\mu_i$ in Each Model. }
\label{fig:PLPpred}
\end{figure}

We here consider to estimate the and price of a spot with floor-area ratio $100\%$ and distance from $1000$m from from the station $i$, namely $\mu_i=\be_0+\be_1+\be_2T_i+\be_3+v_i$ of HNERVF, HNERRD, NER and JN models, and $\mu_i=\exp(\be_0+\be_1+\be_2T_i+\be_3+v_i)$ of GM model.
In Figure \ref{fig:PLPpred}, we provide the predicted values of $\mu_i$ of each model.
From the figure, we can observe that all five models provides relatively similar predicted values, and the predicted values tend to decrease with respect to the area index.
This comes from the effect of $T_i$ since $T_i$ increase as the area index increases.

We finally calculate the mean squared errors (MSE) of predictors.
In JN model, the consistent estimator of MSE cannot be obtained without any knowledge of grouping of areas (stations) as shown in Jiang and Nguyen (2012).
For GM models, the second-order unbiased estimator of MSE is hard to obtain.
Thus, we here consider the MSE estimator of HNERVF, HNERRD and NER models.
We use the analytical estimator given in Theorem \ref{thm:mseest} for HNERVF and NER, and the parametric bootstrap MSE estimator developed in Kubokawa, et al. (2016) is used for HNERRD with 1000 bootstrap replication.
We found that the estimated MSE of HNERRD model is greater than 700 for all areas, while the estimated MSE of HNERVF and NER models are smaller than 20.
The estimated value of shape parameter in dispersion (gamma) distribution in HNERRD is close to $2$, which may inflate the MSE values.
The estimated values of root of MSE (RMSE) of HNERVF and NER models are given in the right panel of Figure \ref{fig:PLP}.
It is revealed that the estimated RMSE of HNERVF is smaller than that of NER in many areas.
In particular, this is true in $37$ areas among $52$ areas.
Especially, in the latter areas, it is observed that the amount of improvement is relatively large.

\begin{table}[!thb]
\caption{The Estimated Regression Coefficients in Each Model}
\label{tab:PLPest}
\begin{center}
\begin{tabular}{cccccccccc}
\toprule
model && $\be_0$ &$\be_1$ &$\be_2$ &$\be_3$ \\
\midrule
HNERVF  && 42.31 & 2.81 & -3.56 & -0.661 \\ 
HNERRD && 37.72 & 3.88 & -3.24 & -0.960 \\
NER && 33.35 & 6.58 & -3.18 & -0.832\\
JN && 37.01 & 3.41 & -2.59 & -3.19\\
GM && 3.63 &  0.168 & -0.122 & -0.039\\
\midrule
\end{tabular}
\end{center}
\end{table}

\section{Concluding Remarks}
In the context of small-area estimation, homogeneous nested error regression models have been extensively studied so far in the literature.
However, some real data sets show heteroscedasticity in variances as pointed out in Jiang and Nguyen (2012).
To extend the traditional homogeneous nested error regression models, Jiang and Nguyen (2012) and Kubokawa, et al. (2016) have proposed heteroscedastic nested error regression models, respectively.
The drawback of the two models is the normality assumption is required for the response values. 
To overcome the problem, we have proposed the structure of unit-level heteroscedastic variances modeled by some covariates and unknown parameters and suggested heteroscedastic nested error regression models without assuming specific underlying distributions.
In terms of the variance modeling with covariates, the generalized linear mixed models are also popular tools, but it requires somewhat strong parametric assumptions.
Therefore, HNERVF model has clear benefits in real application.
Conversely, one drawback of HNERVF is probably the structure of heteroscedastic variances specified by some covariates and unknown parameters, while two heteroscedastic models by Jiang and Nguyen (2012) and Kubokawa, et al. (2016) do not requires such a specific structure.
However, the heteroscedastic variances can be often modeled by some covariates as in the real data application given in Section \ref{sec:app}.

\vspace{0.5cm}\noindent{\bf Acknowledgments.}\\
We appreciate the valuable comments and suggestions from the Associate editors and the reviewer, which lead to the improved version of the paper.
The first author was supported in part by Grant-in-Aid for Scientific Research (15J10076) from Japan Society for the Promotion of Science (JSPS). 
The second author was supported in part by Grant-in-Aid for Scientific Research  (15H01943 and 26330036) from Japan Society for the Promotion of Science.

\vspace{1cm}
\begin{center}
{\bf Appendix}
\end{center}

\vspace{0.5cm}\noindent
{\bf A.1. Proof of Theorem 1.}\ \ \ 
Since $\y_1,\ldots, \y_m$ are mutually independent, the consistency of $\bgah$ follows from the standard argument, so that $\tah^2$ and $\bbeh$ are also consistent.
In what follows, we derive the asymptotic expressions of the estimators.

First we consider the asymptotic approximation of $\tah^2-\tau^2$.
From (\ref{tau}), we obtain
\begin{align}
\tah^2-\tau^2&=\frac1N\sum_{i=1}^m\sum_{j=1}^{n_i}\left\{(y_{ij}-\x_{ij}'\Obbeh)^2-\sih_{ij}^2\right\}-\tau^2\notag\\
&=\frac1N\sum_{i=1}^m\sum_{j=1}^{n_i}\left\{(y_{ij}-\x_{ij}'\bbe)^2-\si_{ij}^2\right\}-\tau^2-\frac1N\sum_{i=1}^m\sum_{j=1}^{n_i}\si_{ij(1)}^2\z_{ij}'(\bgah-\bga)\notag\\
&-\frac2N\sum_{i=1}
^m\sum_{j=1}^{n_i}(y_{ij}-\x_{ij}'\bbe)\x_{ij}'(\Obbeh-\bbe)+o_p(\bgah-\bga)+o_p(\Obbeh-\bbe)\notag\\
&=\frac1m\sum_{i=1}^mu_{1i}-\frac1N\sum_{i=1}^m\sum_{j=1}^{n_i}\si_{ij(1)}^2\z_{ij}'(\bgah-\bga)+o_p(m^{-1/2})+o_p(\bgah-\bga),\label{tau.expand}
\end{align}
where $u_{1i}=mN^{-1}\sum_{j=1}^{n_i}\left\{(y_{ij}-\x_{ij}'\bbe)^2-\siij^2\right\}-\tau^2$ and we used the fact that $\Obbeh-\bbe=O_p(m^{-1/2})$ and $
N^{-1}\sum_{i=1}^m \sum_{j=1}^{n_i}(y_{ij}-\x_{ij}'\bbe)\x_{ij}=O_p(m^{-1/2})$ from the central limit theorem.

For the asymptotic expansion of $\bgah$, remember that the estimator $\bgah$ is given as the solution of the estimating equation
$$
\frac1N\sum_{i=1}^m\sum_{j=1}^{n_i}\left[\left\{y_{ij}-\bary_i-(\x_{ij}-\barx_i)'\Obbeh\right\}^2\z_{ij}-\siij^2(\z_{ij}-2n_i^{-1}\z_{ij}+n_i^{-1}\barz_i)\right]=\0
$$
Using Taylor expansions, we have
\begin{align*}
\0&=\frac1m\sum_{i=1}\u_{2i}-\frac2N\sum_{i=1}^m\sum_{j=1}^{n_i}\left\{y_{ij}-\bary_i-(\x_{ij}-\barx_i)'\bbe\right\}\z_{ij}(\x_{ij}-\barx_i)'(\Obbeh-\bbe)\\
&\ \ \  -\frac1N\sum_{i=1}^m\sum_{j=1}^{n_i}\si_{ij(1)}^2(\z_{ij}-2n_i^{-1}\z_{ij}+n_i^{-1}\barz_i)\z_{ij}'(\bgah-\bga)+o_p(\bgah-\bga)+o_p(m^{-1/2}),
\end{align*}
where 
$$
\u_{2i}=mN^{-1}\sum_{j=1}^{n_i}\left[\left\{y_{ij}-\bary_i-(\x_{ij}-\barx_i)'\bbe\right\}^2\z_{ij}-\siij^2(\z_{ij}-2n_i^{-1}\z_{ij}+n_i^{-1}\barz_i)\right].
$$
From the central limit theorem, it follows that
$$
\frac1N\sum_{i=1}^m\sum_{j=1}^{n_i}\left\{y_{ij}-\bary_i-(\x_{ij}-\barx_i)'\bbe\right\}\z_{ij}(\x_{ij}-\barx_i)'=O_p(m^{-1/2}),
$$
so that the second terms in the expansion formula is $o_p(m^{-1/2})$.
Then we get
$$
\bgah-\bga=\frac{N}{m}\left(\sum_{i=1}^m\sum_{j=1}^{n_i}\si_{ij(1)}^2(\z_{ij}-2n_i^{-1}\z_{ij}+n_i^{-1}\barz_i)\z_{ij}'\right)^{-1}\sum_{i=1}^m\u_{2i}+o_p(\bgah-\bga)+o_p(m^{-1/2}).
$$ 
Under Assumption (A), we have 
$$
\sum_{i=1}^m\sum_{j=1}^{n_i}\si_{ij(1)}^2(\z_{ij}-2n_i^{-1}\z_{ij}+n_i^{-1}\barz_i)\z_{ij}'=O(m).
$$
From the independence of $\y_1,\ldots,\y_m$ and the fact $E(\u_{2i})=\0$, we can use the central limit theorem to show that the leading term in the expansion of $\bgah-\bga$ is $O_p(m^{-1/2})$.
Thus, 
$$
\bgah-\bga=\frac{N}{m}\left(\sum_{i=1}^m\sum_{j=1}^{n_i}\si_{ij(1)}^2(\z_{ij}-2n_i^{-1}\z_{ij}+n_i^{-1}\barz_i)\z_{ij}'\right)^{-1}\sum_{i=1}^m\u_{2i}+o_p(m^{-1/2}).
$$

Using the approximation of $\bgah-\bga$ and $\bgah-\bga=O_p(m^{-1/2})$, we get the asymptotic expression of $\tah^2-\tau^2$ from (\ref{tau.expand}), which establishes the result for $\tah^2$ and $\bgah$.

Finally we consider the asymptotic expansion of $\bbeh-\bbe$.
From the expression in (\ref{bbe}), it follows that
\begin{align*}
\bbeh-\bbe&=\bbet-\bbe+\sum_{s=1}^q\left(\frac{\partial}{\partial\ga_s}\bbet\right)'(\gah_s-\ga)+\left(\frac{\partial}{\partial\tau^2}\bbet\right)'(\tah^2-\tau^2)+o_p(\bgah-\bga)+o_p(\tah^2-\tau^2).
\end{align*}
Since 
$$
\frac{\partial}{\partial\tau^2}\bSi_i=\J_{n_i}, \ \ \ \ \frac{\partial}{\partial\ga_s}\bSi_i=\W_{i(s)}, \ \ s=1,\ldots,q,
$$
for $\W_{i(s)}=\diag(\si_{i1(1)}^2z_{i1s},\ldots,\si_{in_i(1)}^2z_{in_is})$, we have
\begin{equation}\label{bbet.ast}\begin{split}
\frac{\partial}{\partial \tau^2}\bbet&=\left(\X'\bSi^{-1}\X\right)^{-1}\left(\sum_{i=1}^m\X_i'\bSi_i^{-1}\J_{n_i}\bSi_i^{-1}\X_i\right)\left(\bbet_{\tau}^{\ast}-\bbet\right),\\
\frac{\partial}{\partial \ga_s}\bbet&=\left(\X'\bSi^{-1}\X\right)^{-1}\left(\sum_{i=1}^m\X_i'\bSi_i^{-1}\W_{i(s)}\bSi_i^{-1}\X_i\right)\left(\bbet_{\ga_s}^{\ast}-\bbet\right),\ \ s=1\ldots,q,
\end{split}\end{equation}
where
\begin{align*}
\bbet^{\ast}_{\tau}&=\left(\sum_{i=1}^m\X_i'\bSi_i^{-1}\J_{n_i}\bSi_i^{-1}\X_i\right)^{-1}\sum_{i=1}^m\X_i'\bSi_i^{-1}\J_{n_i}\bSi_i^{-1}\y_i,\\
\bbet^{\ast}_{\ga_s}&=\left(\sum_{i=1}^m\X_i'\bSi_i^{-1}\W_{i(s)}\bSi_i^{-1}\X_i\right)^{-1}\sum_{i=1}^m\X_i'\bSi_i^{-1}\W_{i(s)}\bSi_i^{-1}\y_i, \ \ s=1,\ldots,q.
\end{align*}
Under Assumption (A), we have $\bbet^{\ast}_a-\bbe=O_p(m^{-1/2})$ for $a\in \{\tau,\ga_1,\ldots,\ga_q\}$, whereby $\bbet^{\ast}-\bbet=O_p(m^{-1/2})$.
Since $\bgah-\bga=O_p(m^{-1/2})$ and $\tah^2-\tau^2=O_p(m^{-1/2})$ as shown above, we get 
\begin{equation*}
\bbeh-\bbe=\left(\X'\bSi^{-1}\X\right)^{-1}\sum_{i=1}^m\X_i\bSi^{-1}(\y_i-\X_i\bbe)+o_p(m^{-1/2}),
\end{equation*}
which completes the proof.

\vspace{0.5cm}\noindent
{\bf A2. Proof of Corollary \ref{cor:cond}.}\ \ \ 
Let $\bth=(\theta_1,\ldots,\theta_{p+q+1})'=(\bbe',\bga',\tau^2)'$. 
Note that $\psi_i^{\theta_k}, k=1,\ldots,p+q+1$ does not depend on $\y_1,\ldots,\y_{i-1},\y_{i+1},\ldots,\y_{m}$ and that $\y_1,\ldots,\y_m$ are mutually independent.
Then, 
\begin{align*}
\frac{1}{m^2}&E\left[\left(\sum_{j=1}^m\psi_j^{\theta_k}\right)\left(\sum_{j=1}^m\psi_j^{\theta_l}\right)\bigg| \y_i\right]=\frac{1}{m^2}\sum_{j=1,j\neq i}^{m}E\left[\psi_j^{\theta_k}\psi_j^{\theta_l}\right]+\frac{1}{m^2}\psi_i^{\theta_k}\psi_i^{\theta_l}\\
&=\bOm_{kl}+\frac{1}{m^2}\left\{\psi_i^{\theta_k}\psi_i^{\theta_l}-E\left[\psi_i^{\theta_k}\psi_i^{\theta_l}\right]\right\},
\end{align*}
where $\bOm_{kl}$ is the $(k,l)$-element of $\bOm$ and we used the fact that $E[\psi_j^{\theta_k}|\y_i]=E[\psi_j^{\theta_k}]=0$ for $j\neq i$.
Hence, we get the result from the asymptotic approximation of $\bthh$ given in Theorem \ref{thm:asymp}.

\vspace{0.5cm}\noindent
{\bf A3. Proof of Theorem \ref{thm:cond.bias}.}\ \ \ 
We begin by deriving the conditional asymptotic bias of $\bgah$.
Let $\bgat$ be the solution of the equation
$$
\F(\bga; \bbe)\equiv\frac1N\sum_{i=1}^m\sum_{j=1}^{n_i}\left[\left\{y_{ij}-\bary_i-(\x_{ij}-\barx_i)'\bbe\right\}^2\z_{ij}-\siij^2(\z_{ij}-2n_i^{-1}\z_{ij}+n_i^{-1}\barz_i)\right]=\0
$$
with $\siij^2=\si^2(\z_{ij}'\bga)$.
For notational simplicity, we use $\F$ instead of $\F(\bga;\bbe)$ without any confusion and $F_r,r=1,\ldots,q$ denotes the $r$-th component of $\F$, namely $\F=(F_1,\ldots,F_q)'$.
Define the derivatives $\F_{(\a)}$ and $F_{h(\a\b)}$ by
$$
\F_{(\a)}=\frac{\partial\F}{\partial\a'}, \ \ \ \ \  F_{r(\a\b)}=\frac{\partial^2 F_r}{\partial\a\partial\b'}.
$$
It is noted that $F_{h(\bbe\bga)}=0$.
Expanding $\F(\bgah;\Obbeh)=\0$, we obtain
\begin{align*}
\0=\F+\F_{(\bga)}(\bgah-\bga)+\F_{(\bbe)}(\Obbeh-\bbe)+\frac12\t_1+\frac12\t_2+o_p(m^{-1}),
\end{align*}
where $\t_s=(t_{s1},\ldots,t_{sq}),s=1,2$ for
$$
t_{1r}=(\bgah-\bga)'F_{r(\bga\bga)}(\bgah-\bga), \ \ \ \ \ t_{2r}=(\Obbeh-\bbe)'F_{r(\bbe\bbe)}(\Obbeh-\bbe).
$$
It is also noted that
\begin{align*}
\F_{(\bga)}&=-\frac1m\sum_{k=1}^m\sum_{j=1}^{n_k}\si_{kj(1)}^2(\z_{kj}-2n_k^{-1}\z_{kj}+n_k^{-1}\barz_k)\z_{kj}'\\
\F_{(\bbe)}&=-\frac2N\sum_{k=1}^m\sum_{j=1}^{n_k}\left\{y_{kj}-\bary_k-(\x_{kj}-\barx_k)'\bbe\right\}\z_{ij}(\x_{kj}-\barx_k)', 
\end{align*}
so that $\F_{(\bga)}$ is non-stochastic. 
Thus we have
\begin{align*}
E[\bgah-\bga|\y_i]=-(\F_{(\bga)})^{-1}\left\{E[\F(\bga;\bbe)|\y_i]+E\left[\F_{(\bbe)}(\Obbeh-\bbe)\Big|\y_i\right]+\frac12E[\t_1|\y_i]+\frac12E[\t_2|\y_i]\right\}+o_p(m^{-1}).
\end{align*}
In what follows, we shall evaluate the each term in the parenthesis in the above expression. 
For the first term, since $\y_1,\ldots,\y_m$ are mutually independent and $E(\u_{2i})=\0$, we have
$$
E[\F(\bga;\bbe)|\y_i]={1\over m} \u_{2i}.
$$
For evaluation of the second term, we define $\Z_{kr}=\diag(z_{k1r},\ldots,z_{kn_kr})$, where $z_{kjr}$ denotes the $r$-th element of $\z_{kj}$.
Then it follows that
\begin{align*}
E&\left[\F_{r(\bbe)}(\Obbeh-\bbe)\Big|\y_i\right]=-\frac2N\sum_{k=1}^m E\left[(\y_k-\X_k\bbe)'\E_k\Z_{kr}\E_k\X_k(\Obbeh-\bbe)\bigg|\y_i\right]\\
&=-\frac{2}{N}\sum_{k=1,k\neq i}^mE\left[(\y_k-\X_k\bbe)'\E_k\Z_{kr}\E_k\X_k(\Obbeh-\bbe)\Big|\y_i\right]-\frac{2}{N}(\y_i-\X_i\bbe)'\E_i\Z_{ir}\E_i\X_iE\left[\Obbeh-\bbe\Big|\y_i\right].
\end{align*}
Noting that it holds for $\ell=1,\ldots,m$ and $k\neq i$ 
$$
E\left[(\y_\ell-\X_\ell\bbe)(\y_k-\X_k\bbe)'\Big|\y_i\right]=1_{\{\ell=k\}}\bSi_k, \ \ \ \ E[\Obbeh-\bbe|\y_i]=\left(\X'\X\right)^{-1}\X_i'(\y_i-\X_i\bbe),
$$
we have
\begin{align*}
E\bigg[&(\y_k-\X_k\bbe)'\E_k\Z_{kr}\E_k\X_k(\Obbeh-\bbe)\bigg|\y_i\bigg]\\
&=\sum_{\ell=1}^m\tr\left\{\E_k\Z_{kr}\E_k\X_k(\X'\X)^{-1}\X_k'E\left[(\y_\ell-\X_\ell\bbe)(\y_k-\X_k\bbe)'\Big|\y_i\right]\right\}\\
&=\tr\left\{(\X'\X)^{-1}\X_k'\bSi_k\E_k\Z_{kr}\E_k\X_k\right\},
\end{align*}
which is $O(m^{-1})$ and
$$
\frac{1}{N}(\y_i-\X_i\bbe)'\E_k\Z_{kr}\E_k\X_kE\left[\Obbeh-\bbe\Big|\y_i\right]=o_p(m^{-1}).
$$
Thus, we get
\begin{equation}\label{OLS.formula}
E\left[\F_{r(\bbe)}(\Obbeh-\bbe)\Big|\y_i\right]=-\frac2m\sum_{k=1}^m\sum_{j=1}^{n_k} \tr\left\{(\X'\X)^{-1}\X_k'\bSi_k\E_k\Z_{kr}\E_k\X_k\right\}+o_p(m^{-1}),
\end{equation}
where the leading term is $O(m^{-1})$.
For the third and forth terms, note that
$$
F_{r(\bga\bga)}=-\frac1N\sum_{k=1}^m\sum_{j=1}^{n_k}\si_{kj(2)}^2(\z_{kj}-2n_k^{-1}\z_{kj}+n_k^{-1}\barz_k)\z_{kj}'z_{kjr} \ \ \ \ F_{r(\bbe\bbe)}=\frac2N\sum_{k=1}^m \X_k'\E_k\Z_{kr}\E_k\X_k,
$$
which are non-stochastic.
Then for $h=1,\ldots,q$,
\begin{align*}
E[t_{1r}|\y_i]&=-\frac1N\sum_{k=1}^m\sum_{j=1}^{n_k}z_{kjr}\si_{kj(2)}^2(\z_{kj}-2n_k^{-1}\z_{kj}+n_k^{-1}\barz_k)'\bOm_{\bga\bga}\z_{kj}+o_p(m^{-1}),\\
E[t_{2r}|\y_i]&=\frac2N\sum_{k=1}^m\tr\left(\X_k'\E_k\Z_{kr}\E_k\X_k\V_{\rm OLS}\right)+o_p(m^{-1}),
\end{align*}
for $\V_{\rm OLS}=(\X'\X)^{-1}\X'\bSi\X(\X'\X)^{-1}$, where we used Corollary \ref{cor:cond} and
\begin{equation}\label{OLS.cond}
E\left[(\Obbeh-\bbe)(\Obbeh-\bbe)'\big|\y_i\right]=\V_{\rm OLS}+o_p(m^{-1}),
\end{equation}
which follows from the similar argument to the proof of Corollary \ref{cor:cond}.
Thus we obtain
\begin{align*}
E[\t_1|\y_i]&=-\frac1N\sum_{k=1}^m\sum_{j=1}^{n_k}\z_{kj}\si_{kj(2)}^2(\z_{kj}-2n_k^{-1}\z_{kj}+n_k^{-1}\barz_k)'\bOm_{\bga\bga}\z_{kj}+o_p(m^{-1}),\\
E[\t_2|\y_i]&=\frac2N\sum_{k=1}^m\left\{\tr\left(\X_k'\E_k\Z_{kr}\E_k\X_k\V_{\rm OLS}\right)\right\}_r+o_p(m^{-1}),
\end{align*}
where $\{\a_r\}_r$ denotes the $q$-dimensional vector $(a_1,\ldots,a_q)$. 
Therefore, we have established the result for $\bgah$ in (\ref{cond.bias}).

\bigskip\bigskip
We next derive the result for $\tah^2$.
Let 
$$
\tat^2=\frac1N\sum_{k=1}^m\left\{(\y_k-\X_k\bbe)'(\y_k-\X_k\bbe)-\sum_{j=1}^{n_k}\si_{kj}^2\right\}.
$$
Using the Taylor series expansion, we have
\begin{align*}
\tah^2&=\tat^2+\frac{\partial\tat^2}{\partial \bga}(\bgah-\bga)+\frac12(\bgah-\bga)'\left(\frac{\partial^2\tat^2}{\partial\bga\partial\bga'}\right)(\bgah-\bga)\\
&\ \ +\frac{\partial\tat^2}{\partial \bbe}(\Obbeh-\bbe)+\frac12(\Obbeh-\bbe)'\left(\frac{\partial^2\tat^2}{\partial\bbe\partial\bbe'}\right)(\Obbeh-\bbe)+o_p(m^{-1}),
\end{align*}
where we used the fact that $\partial^2\tat^2/\partial\bga\partial\bbe'=0$.
The straight calculation shows that 
$$
\frac{\partial\tat^2}{\partial\bga}=-\frac1N\sum_{k=1}^m\sum_{j=1}^{n_k}\si_{kj(1)}^2\z_{kj}, \ \ \ \ \frac{\partial^2\tat^2}{\partial\bga\partial\bga'}=-\frac1N\sum_{k=1}^m\sum_{j=1}^{n_k}\si_{kj(2)}^2\z_{kj}\z_{kj}',\ \ \ \frac{\partial^2\tat^2}{\partial\bbe\partial\bbe'}=\frac2N\sum_{k=1}^m\X_i'\X_i,
$$
which are non-stochastic.
Thus we obtain
\begin{align*}
E&[\tah^2-\tau^2|\y_i]=E[\tat^2-\tau^2|\y_i]+\left(\frac{\partial\tat^2}{\partial \bga}\right)'E\left[\bgah-\bga|\y_i\right]+\frac12\tr\left\{\left(\frac{\partial^2\tat^2}{\partial\bga\partial\bga'}\right)E\left[(\bgah-\bga)(\bgah-\bga)'\big|\y_i\right]\right\}\\
&+E\left[\left(\frac{\partial\tat^2}{\partial \bbe}\right)'(\Obbeh-\bbe)\bigg|\y_i\right]+\frac12\tr\left\{\left(\frac{\partial^2\tat^2}{\partial\bbe\partial\bbe'}\right)E\left[(\Obbeh-\bbe)(\Obbeh-\bbe)'\big|\y_i\right]\right\}+o_p(m^{-1})\\
&\equiv B_{\tau 1}(\y_i)+B_{\tau 2}(\y_i)+B_{\tau 3}(\y_i)+B_{\tau 4}(\y_i)+B_{\tau 5}(\y_i)+o_p(m^{-1}).
\end{align*}
From the expression of $\tat^2$, it holds that
\begin{align*}
B_{\tau 1}(\y_i)&=\frac1N\sum_{k=1,k\neq i}^{m}n_k\tau^2+\frac1N\left\{(\y_i-\X_i\bbe)'(\y_i-\X_i\bbe)-\sum_{j=1}^{n_i}\siij^2\right\}-\tau^2\\
&=\left(1-\frac{n_i}{N}\right)\tau^2+{1\over m}u_{1i}+\frac{n_i}{N}\tau^2-\tau^2=\frac1mu_{1i},
\end{align*}
for $u_{1i}$ defined in (\ref{u1}).
Also, we immediately have 
$$
B_{\tau 2}(\y_i)=-\frac1N\sum_{k=1}^m\sum_{j=1}^{n_k}\si_{kj(1)}^2\z_{kj}'\b_{\ga}^{(i)}(\y_i)
$$
For evaluation of $B_{\tau 4}(\y_i)$, note that 
$$
\frac{\partial\tat^2}{\partial\bbe}=-\frac{2}{N}\sum_{k=1}^m\X_k'(\y_k-\X_k\bbe).
$$
Similarly to (\ref{OLS.formula}), we get
\begin{align*}
B_{\tau 4}(\y_i)&=-\frac{2}{N}\sum_{k=1}^mE\left[(\y_k-\X_k\bbe)'\X_k(\Obbeh-\bbe)\bigg|\y_i\right]\\
&=-\frac{2}{N}\sum_{k=1}^m\tr\left\{(\X'\X)^{-1}\X_k'\bSi_k\X_k\right\}+o_p(m^{-1}).
\end{align*}
Moreover, Corollary \ref{cor:cond} and (\ref{OLS.cond}) enable us to obtain the expression of $B_{\tau 3}(\y_i)$ and $B_{\tau 5}(\y_i)$, whereby we get
$$
b_{\tau}^{(i)}(\y_i)=m^{-1}u_{1i}-\frac1N\sum_{k=1}^m\sum_{j=1}^{n_k}\si_{kj(1)}^2\z_{kj}'\left\{\b_{\ga}^{(i)}(\y_i)-\b_{\ga}\right\}+b_{\tau},
$$
which completes the proof for $\tah^2$ in (\ref{cond.bias}).

\bigskip
We finally derive the result for $\bbeh$.
By the Taylor series expansion, 
\begin{align*}
\bbeh-\bbe&=\bbet-\bbe+\sum_{s=1}^q\left(\frac{\partial}{\partial\ga_s}\bbet\right)(\gah_s-\ga)+\left(\frac{\partial}{\partial\tau^2}\bbet\right)(\tah^2-\tau^2)+o_p(m^{-1}),
\end{align*}
since 
$$
\left(\frac{\partial\bbet}{\partial\bphi}\right)'(\bphih-\bphi)(\bphih-\bphi)'\left(\frac{\partial\bbet}{\partial\bphi}\right)=o_p(m^{-1}),
$$
from $\partial\bbet/\partial\bphi=O_p(m^{-1/2})$ as shown in the proof of Theorem \ref{thm:asymp}.
From (\ref{bbet.ast}), we have
\begin{align*}
\sum_{s=1}^q&\left(\frac{\partial}{\partial\ga_s}\bbet\right)(\gah_s-\ga_s)\\
&\ \ \ =\left(\X'\bSi^{-1}\X\right)^{-1}\sum_{s=1}^q\left(\sum_{k=1}^m\X_i'\bSi_i^{-1}\W_{i(s)}\bSi_i^{-1}\X_i\right)\left\{\left(\bbet_{\ga_s}^{\ast}-\bbe\right)(\gah_s-\ga_s)-(\bbet-\bbe)(\gah_s-\ga_s)\right\},
\end{align*}
and
$$
\left(\frac{\partial}{\partial\tau^2}\bbet\right)(\tah^2-\tau^2)=\left(\X'\bSi^{-1}\X\right)^{-1}\left(\sum_{k=1}^m\X_k'\bSi_k^{-1}\J_{n_k}\bSi_k^{-1}\X_k\right)\left\{(\bbet_{\tau}^{\ast}-\bbe)(\tah^2-\tau^2)-(\bbet-\bbe)(\tah^2-\tau^2)\right\}.
$$
Let $\bOm_{\beta^{\ast}\ga_s}=E[(\bbet_{\ga_s}^{\ast}-\bbe)(\gah_s-\ga_s)]$ and  $\bOm_{\beta^{\ast}\tau}=E[(\bbet_{\tau}^{\ast}-\bbe)(\tah-\tau)]$.
Then it can be shown that 
$$
E[(\bbet_{\tau}^{\ast}-\bbe)(\tah-\tau)|\y_i]=\bOm_{\beta^{\ast}\ga_s}+o_p(m^{-1}), \ \ \ E[(\bbet_{\ga_s}^{\ast}-\bbe)(\gah_s-\ga_s)|\y_i]=\bOm_{\beta^{\ast}\tau}+o_p(m^{-1}),
$$
which can be proved by the same arguments as in Corollary \ref{cor:cond}.
Thus from Corollary \ref{cor:cond} and the fact that
$$
E\left[\bbet-\bbe|\y_i\right]=\left(\X'\bSi^{-1}\X\right)^{-1}\X_i'\bSi_i^{-1}(\y_i-\X_i\bbe),
$$
we obtain the result for $\bbeh$ in (\ref{cond.bias}).

\vspace{0.5cm}\noindent
{\bf A4. Proof of (\ref{R2.proof}).}\ \ \ 
From the expansion of $\muh_i$, we have
$$
E\left[(\muh_i-\mut_i)^2\right]=E\left[\left\{\left(\frac{\partial\mut_i}{\partial\bth}\right)'(\bthh-\bth)\right\}^2\right]+\frac12 U_1+\frac14U_2,
$$
where
\begin{align*}
U_1
&=E\left[\left(\frac{\partial\mut_i}{\partial\bth}\right)'(\bthh-\bth)(\bthh-\bth)'\left(\frac{\partial^2\mut_i}{\partial\bth\partial\bth'}\Big|_{\bth=\bth^{\ast}}\right)(\bthh-\bth)\right]\\
U_2
&= E\left[\left\{(\bthh-\bth)'\left(\frac{\partial^2\mut_i}{\partial\bth\partial\bth'}\Big|_{\bth=\bth^{\ast}}\right)(\bthh-\bth)\right\}^2\right].
\end{align*}
It is noted that 
$$
U_1=\sum_{j=1}^{p+q+1}\sum_{k=1}^{p+q+1}\sum_{\ell=1}^{p+q+1}
E\left[\left(\frac{\partial\mut_i}{\partial\th_j}\right)\left(\frac{\partial^2\mut_i}{\partial\th_k\partial\th_{\ell}}\Big|_{\bth=\bth^{\ast}}\right)(\thh_j-\th_j)(\thh_k-\th_k)(\thh_{\ell}-\th_{\ell})\right]
\equiv \sum_{j=1}^{p+q+1}\sum_{k=1}^{p+q+1}\sum_{\ell=1}^{p+q+1} U_{1jk\ell},
$$
and
\begin{align}
|U_{1jkl}|
&\leq E\left[\bigg|\left(\frac{\partial\mut_i}{\partial\th_j}\right)\left(\frac{\partial^2\mut_i}{\partial\th_k\partial\th_{\ell}}\Big|_{\bth=\bth^{\ast}}\right)\bigg| \Big|(\thh_j-\th_j)(\thh_k-\th_k)(\thh_{\ell}-\th_{\ell})\Big|\right]\notag \\
&\leq 
E\left[\bigg|\left(\frac{\partial\mut_i}{\partial\th_j}\right)\left(\frac{\partial^2\mut_i}{\partial\th_k\partial\th_{\ell}}\Big|_{\bth=\bth^{\ast}}\right)\bigg|^4\right]^{1/4} 
E\left[\Big|(\thh_j-\th_j)(\thh_k-\th_k)(\thh_{\ell}-\th_{\ell})\Big|^{4/3}\right]^{3/4}
\label{U1}
\end{align}
using Holder's inequality.
Since both $\partial\mut_i/\partial\th_j$ and $\partial^2\mut_i/\partial\th_k\partial\th_{\ell}$ are linear functions of $\y_i$, the first term of (\ref{U1}) is finite under Assumption (A).
Moreover, from Theorem \ref{thm:asymp}, it follows $\sqrt{m}|\thh_j-\th_j|\leq C(\y)$ for some quadratic function of $\y$, so that the second term in (\ref{U1}) is also finite.
Hence, we have $U_1=o(m^{-1})$. 
Similarly, we also obtain $U_2=o(m^{-1})$.
Therefore, using Corollary \ref{cor:cond}, we have
\begin{align*}
E\left[(\muh_i-\mut_i)^2\right]&=E\left[\left\{\left(\frac{\partial\mut_i}{\partial\bth}\right)'(\bthh-\bth)\right\}^2\right]+o(m^{-1})\\
&=\tr\left\{E\left[\left(\frac{\partial\mut_i}{\partial\bth}\right)\left(\frac{\partial\mut_i}{\partial\bth}\right)'E\left((\bthh-\bth)(\bthh-\bth)'\Big | \y_i\right)\right]\right\}+o(m^{-1})\\
&=\tr\left\{E\left[\left(\frac{\partial\mut_i}{\partial\bth}\right)\left(\frac{\partial\mut_i}{\partial\bth}\right)'\bOm+\left(\frac{\partial\mut_i}{\partial\bth}\right)\left(\frac{\partial\mut_i}{\partial\bth}\right)'c(\y_i)o(m^{-1})\right]\right\}+o(m^{-1})\\
&=\tr\left\{E\left[\left(\frac{\partial\mut_i}{\partial\bth}\right)\left(\frac{\partial\mut_i}{\partial\bth}\right)'\right]\bOm\right\}+o(m^{-1})
\end{align*}
since $c(\y_i)$ is fourth-order function of $\y_i$ and $\partial\mut_i/\partial\bth$ is a linear function of $\y_i$, which completes the proof.

\vspace{0.5cm}\noindent
{\bf A5. Derivation of $R_{31i}(\bphi,\bka)$.}\ \ \ 
Since $\y_i$ given $v_i,\bep_i$ is non-stochastic, we have
\begin{align*}
E&\left[\left(\frac{\partial\mut_i}{\partial\bth}\right)'(\bthh-\bth)w_i\right]\\
&=E\left[E\left[\left(\frac{\partial\mut_i}{\partial\bth}\right)'(\bthh-\bth)w_i\bigg| v_i,\bep_i\right]\right]
=E\left[E(\bthh-\bth |\y_i)'\left(\frac{\partial\mut_i}{\partial\bth}\right)w_i\right]\\
&=E\left[\b_{\bbe}^{(i)}(\y_i)'\left(\frac{\partial\mut_i}{\partial\bbe}\right)w_i\right]+E\left[\b_{\bga}^{(i)}(\y_i)'\left(\frac{\partial\mut_i}{\partial\bga}\right)w_i\right]+E\left[b_{\tau}^{(i)}(\y_i)\left(\frac{\partial\mut_i}{\partial\tau}\right)w_i\right]+o(m^{-1})\\
&\equiv R_{31i}(\bphi)+o(m^{-1}).
\end{align*}
It is noted that $E(w_i)=0$ and 
\begin{equation}\label{iden}
E\left[(y_{ij}-\x_{ij}'\bbe)w_i\right]=E\left[(v_i+\ep_{ij})w_i\right]=\left(\sum_{j=1}^{n_i}\la_{ij}-1\right)\tau^2+\sum_{j=1}^{n_i}\la_{ij}\siij^2=0.
\end{equation}
Using the expression (\ref{cond.bias}) and (\ref{mu.deriv}), it follows that
\begin{align*}
E\left[\b_{\bbe}^{(i)}(\y_i)'\left(\frac{\partial\mut_i}{\partial\bbe}\right)w_i\right]&=\left(\c_i-\sum_{j=1}^{n_i}\la_{ij}\x_{ij}\right)'\left(\X'\bSi^{-1}\X\right)^{-1}\X_i'\bSi_i^{-1}E\big[(\y_i-\X_i\bbe)w_i\big]=0\\
E\left[\b_{\bga}^{(i)}(\y_i)'\left(\frac{\partial\mut_i}{\partial\bga}\right)w_i\right]&=\eta_i^{-2}\sum_{j=1}^{n_i}\siij^{-2}\bde_{ij}'\left(\sum_{k=1}^{m}\sum_{h=1}^{n_k}\si_{kh(1)}^2\z_{kh}\z_{kh}'\right)^{-1}\M_{2ij}(\bphi,\bka)\\
E\left[b_{\tau}^{(i)}(\y_i)\left(\frac{\partial\mut_i}{\partial\tau}\right)w_i\right]&=m^{-1}\eta_i^{-2}\sum_{j=1}^{n_i}\siij^{-2}\bigg\{M_{1ij}(\bphi,\bka)-\T_1(\bga)'\T_2(\bga)\M_{2ij}(\bphi,\bka)\bigg\},
\end{align*}
where
$$
\M_{2ij}(\bphi,\bka)=E\left[\u_{2i}(y_{ij}-\x_{ij}'\bbe)w_i\right], \ \ \ \  M_{1ij}(\bphi,\bka)=E\left[u_{1i}(y_{ij}-\x_{ij}'\bbe)w_i\right].
$$

\medskip
To evaluate $M_{1ij}$ and $\M_{2ij}$, we first prove the following result for fixed $j,k,\ell\in \{1,\ldots,n_i\}$.
\begin{equation}\label{iden2}
\begin{split}
E\big[(v_i+\ep_{ij})(v_i+\ep_{ik})(v_i+\ep_{i\ell})w_i\big]&=\tau^2\eta_i^{-1}\bigg[\tau^2(3-\kappa_v)+\kappa_{\ep}\siij^21_{\{j=k=\ell\}}+\siij^2(1_{\{j=k\neq \ell\}}-1_{\{j=k\}})\\
&\ \ \ \ \ +\siij^2(1_{\{j=\ell\neq k\}}-1_{\{j=\ell\}})+\si_{ik}^2(1_{\{k=\ell\neq j\}}-1_{\{k=\ell\}})\bigg].
\end{split}\end{equation}
To show (\ref{iden2}), we note that the left side can be rewritten as
\begin{equation}\label{iden3}
-\eta_i^{-1}E\left[(v_i+\ep_{ij})(v_i+\ep_{ik})(v_i+\ep_{i\ell})v_i\right]+\sum_{h=1}^{n_i}\la_{ih}E\left[(v_i+\ep_{ij})(v_i+\ep_{ik})(v_i+\ep_{i\ell})\ep_{ih}\right]
\end{equation}
from the definition of $w_i$.
Using the fact that $\ep_{i1},\ldots,\ep_{in_i}$ and $v_i$ are independent, the first term in (\ref{iden3}) is calculated as
\begin{align*}
E\left[v_i^4+(\ep_{ij}\ep_{ik}+\ep_{ij}\ep_{i\ell}+\ep_{ik}\ep_{i\ell})v_i^2\right]=\ka_v\tau^4+\tau^2\left(\siij^21_{\{j=k\}}+\siij^21_{\{j=\ell\}}+\si_{ik}^21_{\{k=\ell\}}\right).
\end{align*}
Moreover, we have
\begin{align*}
E&\left[(v_i+\ep_{ij})(v_i+\ep_{ik})(v_i+\ep_{i\ell})\ep_{ih}\right]=E\left[\ep_{ih}(\ep_{ij}+\ep_{i\ell}+\ep_{ik})v_i^2+\ep_{ij}\ep_{ik}\ep_{i\ell}\ep_{ih}\right]\\
&=\tau^2\si_{ih}^2\left(1_{\{h=j\}}+1_{\{h=k\}}+1_{\{h=\ell\}}\right)+\ka_{\ep}\siih^41_{\{j=k=\ell=h\}}\\
& \ \ \ \ +\siih^2\left(\siij^21_{\{j=k\neq \ell=h\}}+\siij^21_{\{j=\ell\neq k=h\}}+\si_{ik}^21_{\{j=h\neq k=\ell\}}\right),
\end{align*}
whereby the second term in (\ref{iden3}) can be calculated as
$$
\tau^2\eta_i^{-1}\left[3\tau^2+\ka_{\ep}\siij^21_{\{j=k=\ell\}}+\siij^21_{\{j=k\neq\ell\}}+\siij^21_{\{j=\ell\neq k\}}+\si_{ik}^21_{\{k=\ell\neq j\}}\right],
$$
where we used the expression $\la_{ih}=\tau^2\eta_i^{-1}\si_{ih}^{-2}$.
Then we established the result (\ref{iden2}).
From (\ref{iden2}), we immediately have
\begin{align*}
\sum_{\ell=1}^{n_i}E\big[(v_i+\ep_{ij})(v_i+\ep_{ik})(v_i+\ep_{i\ell})w_i\big]&=\tau^2\eta_i^{-1}\left[n_i\tau^2(3-\ka_v)+\siij^2(\ka_{\ep}-3)1_{\{j=k\}}\right]\\
&=E\big[(v_i+\ep_{ij})(v_i+\ep_{ik})^2w_i\big].
\end{align*}

Now, we return to the evaluation of $M_{1ij}$ and $\M_{2ij}$.
It follows that
\begin{align*}
M_{1ij}(\bphi,\bka)&=\frac{m}{N}\sum_{h=1}^{n_i}E\left[(y_{ih}-\x_{ih}'\bbe)^2(y_{ij}-\x_{ij}'\bbe)w_i\right]\\
&=mN^{-1}\eta_i^{-1}\tau^2\Big\{n_i\tau^2(3-\ka_v)+\siij^2(\ka_{\ep}-3)\Big\}
\end{align*}
and
\begin{align*}
\M_{2ij}(\bphi,\bka)&=\frac{m}{N}\sum_{h=1}^{n_i}\z_{ih}E\left[\{v_i+\ep_{ih}-(v_i+\barep_i)\}^2(v_i+\ep_{ij})w_i\right]\\
&=\frac{m}{N}\sum_{h=1}^{n_i}\z_{ih}\bigg\{E\left[(v_i+\ep_{ih})^2(v_i+\ep_{ij})w_i\right]-2n_i^{-1}\sum_{k=1}^{n_i}E\left[(v_i+\ep_{ij})(v_i+\ep_{ik})(v_i+\ep_{ih})w_i\right]\\
&\ \ \ \ +n_i^{-2}\sum_{k=1}^{n_i}\sum_{\ell=1}^{n_i}E\left[(v_i+\ep_{ij})(v_i+\ep_{ik})(v_i+\ep_{i\ell})w_i\right]\bigg\}.
\end{align*}
Using the identity given in (\ref{iden2}), we have
\begin{align*}
\M_{2ij}(\bphi,\bka)&=mN^{-1}\tau^2\eta_i^{-1}\sum_{h=1}^{n_i}\z_{ih}\Big\{\siij^2(\ka_{\ep}-3)(1_{\{j=h\}}-2n_i^{-1}1_{\{j=h\}}+n_i^{-2})\Big\}\\
&=mN^{-1}\tau^2\eta_i^{-1}n_i^{-2}(n_i-1)^2(\ka_{\ep}-3)\siij^2\z_{ij},
\end{align*}
which completes the result in (\ref{R31}).

\vspace{0.5cm}\noindent
{\bf A6. Evaluation of $R_{32i}(\bphi)$.}\ \ \ 
Since $\y_i$ given $v_i$ and $\bep_i$ is non-stochastic, we have
\begin{align*}
R_{32i}(\bphi)&=\frac12E\left[(\bthh-\bth)'\left(\frac{\partial^2\mut_i}{\partial\bth\partial\bth'}\Big|_{\bth=\bth^{\ast}}\right)(\bthh-\bth)w_i\right]=\frac12E\left[E\left[(\bthh-\bth)'\left(\frac{\partial^2\mut_i}{\partial\bth\partial\bth'}\Big|_{\bth=\bth^{\ast}}\right)(\bthh-\bth)w_i\bigg| v_i,\bep_i\right]\right]\\
&=\frac12\tr\left\{\bOm E\left[\left(\frac{\partial^2\mut_i}{\partial\bth\partial\bth'}\Big|_{\bth=\bth^{\ast}}\right)w_i\right]\right\}+o(m^{-1})E\left[\tr\left\{c(\y_i)\left(\frac{\partial^2\mut_i}{\partial\bth\partial\bth'}\Big|_{\bth=\bth^{\ast}}\right)\right\}w_i\right],
\end{align*}
where we used Corollary \ref{cor:cond} in the last equation.
Note that 
\begin{equation}\label{exp2}
\frac{\partial^2\mut_i}{\partial\bth\partial\bth'}\Big|_{\bth=\bth^{\ast}}=\frac{\partial^2\mut_i}{\partial\bth\partial\bth'}+\sum_{k=1}^{p+q+1}(\th_k^{\ast}-\th_k)\left(\frac{\partial^3\mut_i}{\partial\bth\partial\bth'\partial\th_k}\Big|_{\th_k=\th_k^{\ast\ast}}\right),
\end{equation}
where $\th_k^{\ast\ast}$ is an intermediate value between $\th_k^{\ast}$ and $\th_k$.
Further note that the third order partial derivatives of $\mut_i$ is a linear function of $\y_i$, so that the second term of $R_{32i}$ is $o(m^{-1})$.
Similarly, it follows that 
$$
E\left[\left(\frac{\partial^2\mut_i}{\partial\bth\partial\bth'}\Big|_{\bth=\bth^{\ast}}\right)w_i\right]=E\left[\left(\frac{\partial^2\mut_i}{\partial\bth\partial\bth'}\right)w_i\right]+o(1)=o(1),
$$
since the second order partial derivatives of $\mut_i$ is a linear function of $y_{ij}-\x_{ij}'\bbe$ and the identity (\ref{iden}).
Therefore, we finally get $R_{32i}(\bphi)=o(m^{-1})$.


\end{document}